\documentclass[fleqn,usenatbib]{mnras}

\usepackage{newtxtext,newtxmath}

\usepackage[T1]{fontenc}

\DeclareRobustCommand{\VAN}[3]{#2}
\let\VANthebibliography\thebibliography
\def\thebibliography{\DeclareRobustCommand{\VAN}[3]{##3}\VANthebibliography}

\usepackage{graphicx}	
\usepackage{amsmath}	

\newcommand{\hii}{H~{\sc ii}}

\newcommand{\nh}{N$_{\mathrm{HI}}$}
\newcommand{\nhstop}{N$_{\rm{HI,stop}}$}
\newcommand{\fesc}{f$_{\rm{esc}}$}

\newcommand{\Ndot}{$\dot{N}_{\rm{ion}}$}
\newcommand{\xion}{$\xi_{\rm{ion}}$}

\title[The impact of nebular LyC on escape fractions]{The impact of nebular Lyman-Continuum on ionising photons budget and escape fractions from galaxies}

\author[C. Simmonds et al.]{
C. Simmonds,$^{1,2}$\thanks{E-mail: cs2210@cam.ac.uk}
A. Verhamme,$^{3}$
A. K. Inoue,$^{4,5}$
H. Katz,$^{6}$
T. Garel$^{3}$
and S. De Barros$^{3}$
\\
$^{1}$The Kavli Institute for Cosmology (KICC), University of Cambridge, Madingley Road, Cambridge, CB3 0HA, UK\\
$^{2}$Cavendish Laboratory, University of Cambridge, 19 JJ Thomson Avenue, Cambridge, CB3 0HE, UK\\
$^{3}$Observatoire de Genève, Université de Genève, Chemin Pegasi 51, 1290 Versoix, Switzerland\\
$^{4}$Department of Physics, School of Advanced Science and Engineering, Faculty of Science and Engineering, Waseda University, 3-4-1 Okubo, \\Shinjuku, 169-8555 Tokyo, Japan\\
$^{5}$Waseda Research Institute for Science and Engineering, Faculty of Science and Engineering, Waseda University, 3-4-1 Okubo, Shinjuku,\\ 169-8555 Tokyo, Japan\\
$^{6}$Department of Astronomy \& Astrophysics, University of Chicago, 5640 S Ellis Avenue, Chicago, IL 60637, USA\\
}

\date{Accepted XXX. Received YYY; in original form ZZZ}

\begin{document}
\label{firstpage}
\pagerange{\pageref{firstpage}--\pageref{lastpage}}
\maketitle

\begin{abstract}
Several Lyman Continuum (LyC) emitters have been detected so far, but their observed ionising spectra sometimes differ from attenuated stellar spectra predicted by stellar population synthesis modelling. This discrepancy may be due to a significant contribution of LyC nebular emission. We aim to quantify the importance this emission in LyC leakers: its contribution to the ionising photons budget, and to measurements of LyC escape. To estimate the nebular contribution to the LyC spectra of galaxies, we run photoionisation models with \textsc{Cloudy} for a range of BPASS templates, varying the column density of the surrounding gas, from density-bounded (log(\nhstop/cm$^{-2}$)=16) to ionisation-bounded (log(\nhstop/cm$^{-2}$)=19) regimes. In the limits of very optically thin (\fesc\/ = 1), or thick configurations (\fesc\/ = 0), there is no nebular contribution to the emergent LyC spectra. This contribution matters only at intermediate LyC opacities ($0 <$ \fesc\/ $< 1$), where it alters the shape of the LyC spectrum chromatically, so that escape fractions estimates are highly sensitive to the wavelength range over which they are calculated. We propose a formula to estimate integrated escape fractions using f$_{\lambda 700}$/f$_{\lambda 1100}$ flux ratios, since this wavelength range is not affected by nebular emission. Regarding simulations, the boost of hydrogen ionising photons escaping galaxies is inversely proportional to the stellar escape fractions, but since typical simulated escape fractions are low, LyC photons escape is important. Nebular LyC is a non-negligible additional source of ionising photons from galaxies, which contribution has been overlooked so far in observations and in cosmic reionisation simulations.
\end{abstract}




\section{Introduction}
The Epoch of Reionisation (EoR) describes the period when the Universe transitioned from being predominantly neutral to ionised by redshift $z \sim 6$ \citep{Becker2001,Fan2006,Yang2020}, with some studies favouring a later reionisation at $z \sim 5.3$ \citep{Kulkarni2019,Bosman2022}. Although there is some debate regarding the origin of the radiation responsible for reionising the Universe, star-forming galaxies are likely the primary contributors of ionising photons \citep{Hassan2018,Rosdahl2018,Trebitsch2020,Chisholm2022}. For galaxies to reionise the Universe, there are two basic conditions that must be fulfilled: (1) these galaxies need to produce ionising photons (E $>$ 13.6 eV or $\lambda < 912$ \AA, also called Lyman Continuum or LyC), and (2) these photons must escape the interstellar medium (ISM) to ionise the intergalactic medium (IGM). The ratio between the photons that escape to the ones that are produced defines the \textsl{escape fraction} (\fesc\/). A galaxy for which hydrogen ionising radiation is observed to escape is called a LyC leaker. 

Directly observing LyC escaping from galaxies is a challenging task due to a combination of physical effects and instrumental limitations, depending on the redshift of the sources. Ground-based instruments are strictly limited by the absorption of all UV radiation by the atmosphere for galaxies with redshifts up to $z \sim 2.7$. This is a physical barrier that can only be overcome by using space-based instruments. Moreover, as redshift increases, the attenuation from the neutral IGM in the line-of-sight (LOS) becomes increasingly important, with virtually no transparent LOS for LyC above $z \sim 5$ \citep{Madau1995,Inoue2014}. To date, a few dozen LyC leakers have been observed both locally \citep[e.g.][]{Leitet2013,Borthakur2014,Izotov2016b,Leitherer2016,Izotov2018b,Wang2019,Izotov2021,Flury2022a} and at $z \gtrsim 2$ \citep[e.g.][]{Mostardi2015,DeBarros2016,Vanzella2016,Bian2017,Steidel2018,Vanzella2018,Fletcher2019,Rivera-Thorsen2019,Ji2020,Marques-Chaves2021,Marques-Chaves2022,Kerutt2023}. In addition, a few LyC leaker candidates have been detected at intermediate redshifts using the Indian space telescope AstroSat, for example at $z \simeq 1.4$ \citep[AUDFs01;][]{Saha2020}. 

Most of the LyC detections in the literature have been made by observing fluxes at rest-frame wavelengths close to the Lyman break ($\lambda \sim 900$ \AA\/). In this spectral region there is a prominent feature in the nebular spectrum that arises from the recombination of ionised atoms direct to the ground state, followed by the emission of a hydrogen-ionising photon. \cite{Inoue2010} and \cite{Inoue2011} discuss the importance and detectability of the free-bound nebular continuum from LyC leakers with intermediate LyC opacities, corresponding to neutral hydrogen column densities of \nh\ $ = 10^{17} - 10^{18}$ cm$^{-2}$. In this regime, the LyC flux just below the Lyman limit is dominated by this nebular emission. As a result, an excess flux is observed around rest-frame wavelengths of $\lambda \sim 900$ \AA\/ (dubbed Lyman bump). To quantify the effect the nebular emission has on \fesc\/, in this work we use the version 17.01 of the photoionisation code \textsc{Cloudy} \citep{Ferland2017}, combined with stellar population models from the Binary Population and Spectral Synthesis version 2.2.1 \citep[BPASS;][]{Eldridge2017}. \\

In this article, we aim to shed light on the nuances of the \fesc\ definitions, in an effort to compile and clarify the different definitions provided in the literature. We also discuss the impact of the nebular emission (which arises from the gas in the nebula around the ionising sources) on the shape of the observed spectra \citep[Lyman bump;][]{Inoue2010,Inoue2011} and consequently, on \fesc\/. \S~\ref{section:fesc} defines \fesc\ and how this term is used throughout the literature, as well as the uncertainties and assumptions associated with it. \S~\ref{section:data} presents the models and observations used in this work. Followed by our results in \S~\ref{section:results} and their discussion in \S~\ref{section:discussion}. Finally, a small summary and the main conclusions can be found in \S~\ref{section:conclusions}.

\section{Escape fraction definitions from literature}
\label{section:fesc}

Broadly, \fesc\ can refer to a cosmic or a galactic-scale quantity, depending on if we are interested in the reionisation of the Universe or if we are rather concerned with individual galaxy estimations. It refers sometimes to the wavelength integrated quantity, or is computed over a specific wavelength range. It can be an angle-averaged value, or computed along a given line of sight. 

\begin{figure}
   \centering
   \includegraphics[width=\columnwidth]{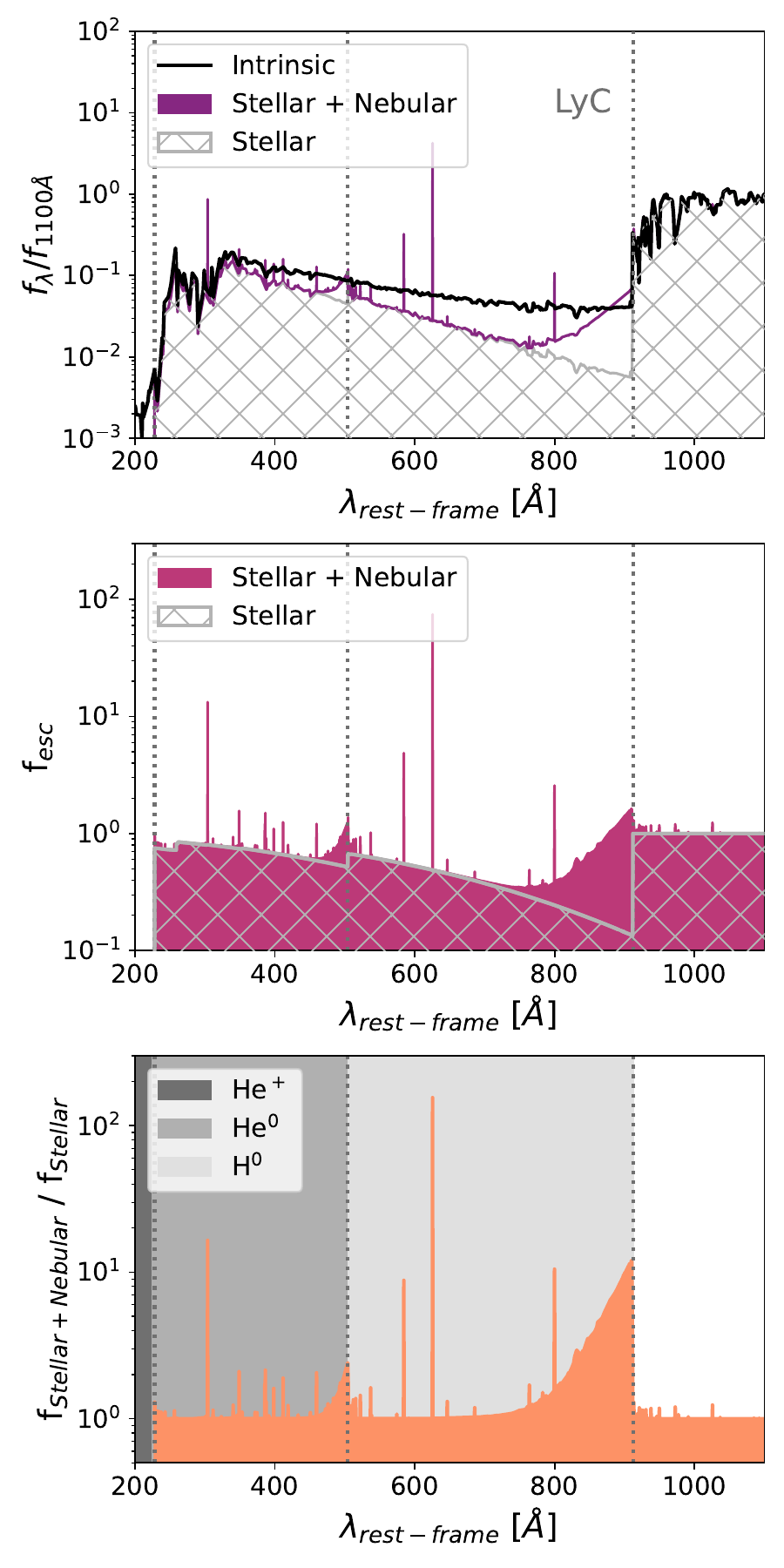}
    \caption{Effect of the inclusion of diffuse fields in the output \textsc{Cloudy} spectra for a 5 Myr old \textsc{Cloudy} model with a constant star formation, a metallicity of Z = 0.006 (fiducial model), and log(\nhstop\/)/cm$^{-2}$ = 17.5. \textsl{Top panel:} intrinsic (black curve), transmitted (grey hatched area) and net transmitted (filled purple area) spectra. The net transmitted spectrum includes the nebular emission. Note the difference the inclusion of the nebular emission makes around the Lyman break at 912 \AA\/. \textsl{Middle panel:} escape fraction as a function of wavelength with (pink shaded area) and without (grey hatched area) the inclusion of the diffuse fields. \textsl{Bottom panel:} ratio between the flux densities with and without diffuse emission, as a function of wavelength. The grey shaded areas show the regions where hydrogen and helium are ionised, as indicated in the legend. It can be seen that there is an excess emission in the vicinity of the ionisation potentials, the strongest being located close to the Lyman limit.}
              \label{fig:diffuse_effect}
   \end{figure}
   
\subsection{Cosmic escape fraction}
The cosmic escape fraction, hereafter f$_{\rm esc, cosmic}$, indicates the fraction of the total ionising photons produced at a given time, by all sources in the Universe, that escape into the IGM. The evolution of this global quantity with redshift helps us understand the history of reionisation of the Universe. The Universe remains ionised if the recombination rate of intergalactic hydrogen is lower than the emissivity of ionising photons from galaxies into the IGM, \Ndot. This is illustrated in the models from \cite{Madau1999}, which provide redshift-dependent curves indicating the \Ndot\ needed in order to maintain hydrogen ionisation in the IGM.  This emissivity is usually parameterised as follows \citep{Madau1999}:
\begin{equation}
    \label{eq:global_fesc}
    \dot{N}_{\rm{ion}} = \text{f}_{\rm esc, cosmic} \times \xi_{\rm{ion}} \times \rho_{\rm{UV}}
\end{equation}

\noindent where \Ndot\ is in units of photon s$^{-1}$ Mpc$^{-3}$, \xion\ is in units of Hz erg$^{-1}$, and $\rho_{\rm{UV}}$ in units of erg s$^{-1}$ Hz$^{-1}$ Mpc$^{-3}$. 
This equation defines the cosmic escape fraction. The ionising photon production efficiency, \xion\/, describes the amount of ionising photons being produced over the non-ionising UV continuum. Promisingly, it has been observed to increase with redshift \citep[e.g. ][]{Stefanon2022,Prieto-Lyon2023,Rinaldi2023,Simmonds2023b}, potentially reducing the required average escape fractions needed in order for galaxies to reionise the Universe \citep[][]{Simmonds2023b}. Finally, a total rest-frame UV luminosity density function, $\rho_{\rm{UV}}$, has to be assumed \citep[constrained by observations, see e.g. ][and references within]{Atek2014,Atek2018,Bouwens2021,Bouwens2022}, which indicates how many objects per unit volume of a certain UV luminosity exist as a function of redshift. 

\begin{figure*}
   \centering
   \includegraphics[width=0.66\textwidth,trim={0 0 0 0}]{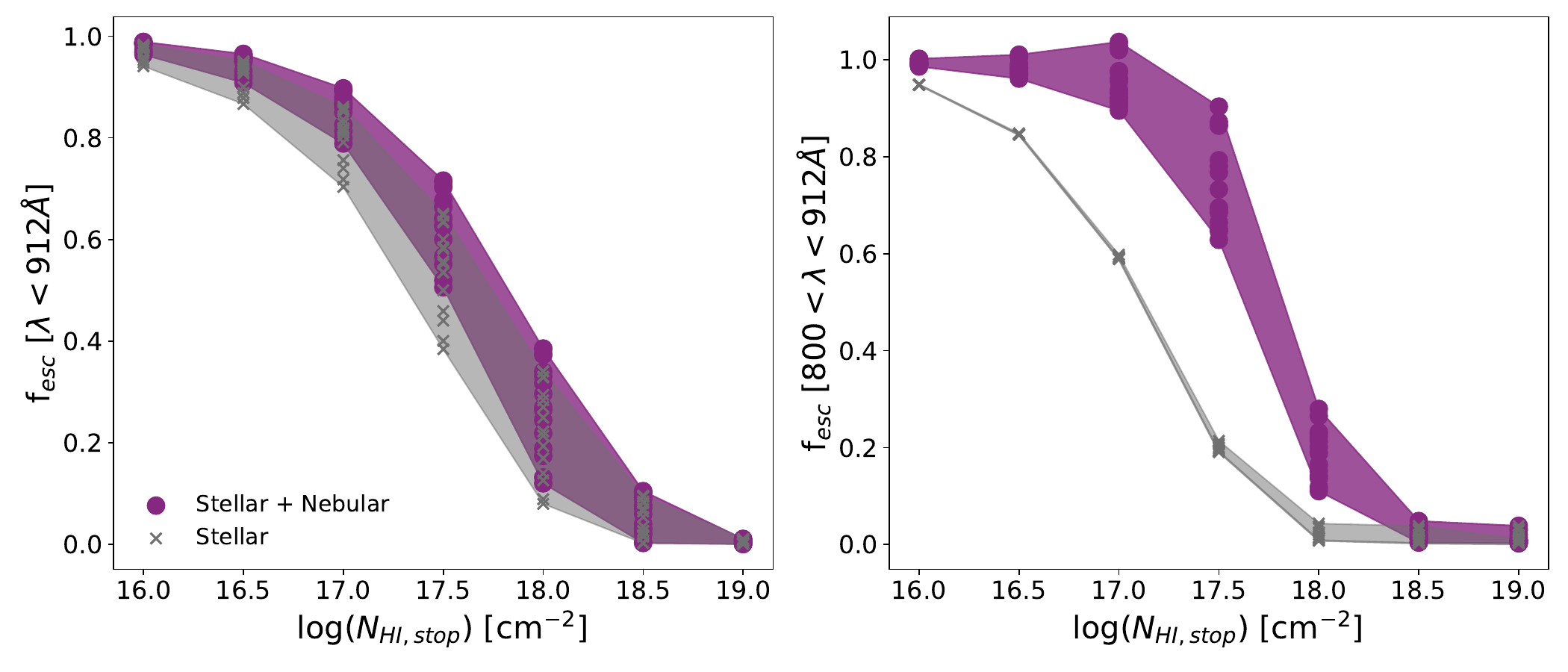}
       \includegraphics[width=0.33\textwidth]{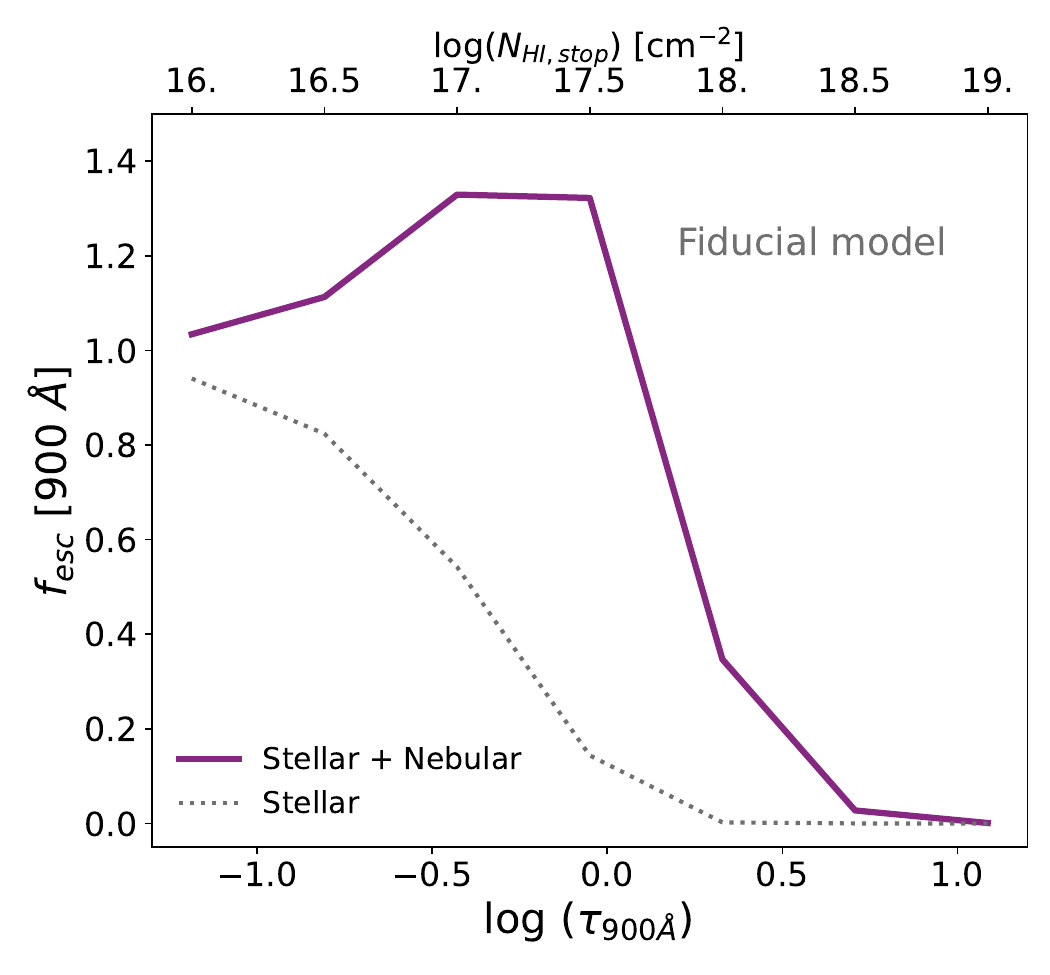}
    \caption{Effect of the LyC nebular emission on escape fractions, when varying \nhstop, the stopping criteria set for the \textsc{Cloudy} models in order to simulate a density bounded region. 
    \textsl{Left panel:} escape fractions integrated in the range $\lambda = 0 - 912$ \AA\/, the vertical spread is driven by the ages and metallicities of the stellar population models. \textsl{Middle panel:} escape fractions integrated in the range $\lambda = 800 - 912$ \AA\/. \textsl{Right panel:} monochromatic escape fractions at 900 \AA\/, where \nhstop has a maximum effect for the fiducial model. We note that even at high \nhstop\/ (i.e. $10^{18}$ cm$^{-2}$), some ionising photons are able to escape.  We also draw attention to the fact that the nebular emission produces only a slight increase in \fesc\ when the full LyC region is integrated, however, a more pronounced effect is produced when the region close to the Lyman limit is considered instead. The ionising spectra of LyC emitters are usually observed in this wavelength range, and therefore, special care must be taken when estimating escape fractions.}
              \label{fig:Lybump}
   \end{figure*}

There are two main sources of uncertainties in Equation~\ref{eq:global_fesc}. (1) The shape of $\rho_{\rm{UV}}$ is constrained by observations, which are biased towards brighter objects. Extensive work has been done to characterise the faint-end slope of $\rho_{\rm{UV}}$, but some uncertainty remains in this regime \citep[][]{Atek2018,Bouwens2023,Thai2023,Harikane2024}. Moreover, the inferred $\rho_{\rm{UV}}$ depends on the sampled volume estimation, which depends on the selection and completeness of the survey used. (2) Once $\rho_{\rm{UV}}$ has been determined, the UV luminosity must be converted to an ionising photon budget through the parameter \xion\/, estimated from stellar populations spectra. Among other parameters, these models are age, metallicity, IMF, binarity, and star-formation history dependent \citep[][]{Leitherer1999,Schaerer2003,Eldridge2017}.   

This parametrisation in itself also implies/contains strong assumptions on the sources of cosmic reionisation: most of the LyC sources accounted for are \emph{observed} star-forming galaxies, but, even if massive stars are most probably the dominant sources of ionising radiation in the Universe, we know that we miss intrinsic LyC photons from obscured star-formation when only counting UV light \citep{Mitsuhashi2023}. Then, quasars and/or faint Active Galactic Nuclei (AGNs) may also contribute to the global ionising photons budget \citep{Eide2020, Trebitsch2021}, and are not taken into account in this formalism. Finally, ionising photons from the gas itself, either collisionally ionised by shocks \citep{Eide2020}, or cosmic rays, or photo-ionised by stars -- which is the core of this study-- are not accounted for. 

When only the un-obscured star-formation is taken into account, from this parametrisation, the cosmic escape fraction has to be of the order 10-20\% at $z\sim6$ to explain reionisation \citep{Inoue2006,Robertson2013,Robertson2015}, although this threshold can be lowered under certain conditions to $< 5$\% \citep[see for example, ][]{Finkelstein2019}. This value may be considered as an upper limit, assuming that there are other probable sources of ionising radiation contributing to the total budget but not much to the escaping budget, although very difficult to quantify \citep[but this may change in the near future, thanks to JWST,][]{barrufet2023, Nelson2023}. Indeed, on the simulation side, the cosmic escape fraction is predicted to be smaller than the canonical 10-20\% required from equation~\ref{eq:global_fesc}: 5\% at z$\sim 6$ for the  THESAN project \citep{Yeh2023}, as well as for DUSTiER \citep{Lewis2023} and OBELISK \citep{Trebitsch2021}, and 2-6\% at z$\sim 6$ from the SPHINX suite of cosmological simulations \citep{Rosdahl2018, Rosdahl2022}.

\subsection{Galactic-scale escape fractions}
Galactic-scale escape fractions, \fesc\/$_{\rm{,gal}}$ are estimated for individual galaxies. As in the cosmic case, the key components are the amount of ionising photons that escape the galaxy versus the ones that are being produced intrinsically. The studies on this topic can be divided into results obtained through numerical simulations or through observations, and can be global (averaged in all directions of view), wavelength-integrated, or calculated along specific directions of observation and over specific wavelength ranges. 

Simulations have the advantage of having a complete 4$\pi$ view of the galaxy, where the intrinsic spectra is completely known and so, \fesc\/$_{\rm{,gal}}$ can be calculated in a straightforward manner, with LyC fluxes taken at specific wavelengths or rather integrated in the full $\lambda = 0 - 912$ \AA\ range; either globally \citep{Yajima2011, Paardekooper2015,Xu2016,Ma2020,Rosdahl2022, Maji2022, Katz2023, Seeyave2023,Yeh2023} or in particular lines-of-sight \citep{Mauerhofer2021, Katz2023}. One remarkable result from \cite{Mauerhofer2021} (see figure 12) is that line-of-sight escape fractions are not distributed around the global escape fraction, \fesc\/$_{\rm{,gal,4\pi}}$, but at each time step the galaxy is non-leaking along most directions of observations. There is a tail of non-zero directional escape fractions, \fesc\/$_{\rm{,gal,LoS}}$, in a small fraction of directions, however, so that the median of the distribution of directional escape fractions is different from the global escape fraction. Most radiation hydrodynamics simulations consider stars as the only sources of ionising radiation, a few take into account AGN contribution \citep[c.f.][]{Trebitsch2021,Trebitsch2023}, but in both cases the escape fractions are calculated as exp$(-\tau)$\footnote{Dust absorption and scattering of ionising photons is also modelled in these studies, and shown to have a negligible effect on \fesc\/ [but see also \citet{Inoue2001AJ} and \citet{Inoue2001ApJ}]}, and the possible contribution of nebular LyC emission is neglected. 

In comparison, observations are always dependent on the viewing angle and the rest-frame wavelengths probed, \fesc\/$_{\rm{,gal,LoS,\Delta\lambda}}$. Most of the observations probe LyC around rest-frame $\lambda = 800 - 900$ \AA\/ \citep[e.g. ][]{Steidel2018, Izotov2021, Flury2022a}, with some exceptions. For example, LyC was observed in AUDFs01 at rest-frame $\lambda \sim 600$ \AA\/ \citep{Saha2020}.
 When some LyC flux is detected from a galaxy, we need to estimate the intrinsic LyC luminosity, over the same wavelength range, in order to calculate \fesc\/$_{\rm{,gal,LoS,\Delta\lambda}}$. As detailed in \cite{Flury2022a}, there are two ways to determine \fesc\/$_{\rm{,gal,LoS,\Delta\lambda}}$. (1) If H$\alpha$ or another recombination line of hydrogen has been observed, we can convert its (dust corrected) flux into a luminosity of hydrogen ionising photons that did not escape, add it to the observed LyC luminosity, i.e. to the luminosity of hydrogen ionising photons that escaped, to get an estimate of the intrinsic hydrogen ionising photons budget. Uncertainties with this approach come from the dust correction factor, and the conversion factor between the strength of the line and the corresponding intrinsic LyC luminosity. Using H$\alpha$ to estimate the LyC luminosity that did not escape the galaxy assumes that LyC photons absorbed by dust, HeI and HeII are negligible. 
 (2) Another possibility is to fit the (dust corrected) UV non-ionising stellar continuum with a library of Spectral Energy Distribution (SED) models to estimate the intrinsic LyC photons budget over the same wavelength range.
 In this case, dust attenuation also introduces uncertainties on the recovered UV stellar continuum, and the LyC estimates rely on stellar library models (see section~\ref{section:LyC_dependence_on_intrinsic_spectra} below). The assumption of this method is that stars are the only sources of ionising radiation.
 Another factor that complicates the estimation of \fesc\ for high-redshift LyC detections is the stochastic neutral IGM that increases with redshift \citep{Inoue2014}. 


Although very intuitive in concept, the several escape fraction estimates in the literature always come with a set of assumptions and limitations that need to be considered to make meaningful comparisons between studies. 
To conclude this discussion on the concept of escape fractions, we would like to mention yet another complication: the recombining gas emits a significant quantity of ionising photons, e.g. through free-bound emission, and along leaking lines of sights, some of these may escape. Since this gas has most probably been photoionised by the stars, should this radiation be counted in the numerator or the denominator of the escape fraction? as a (secondary) source term, or as escaping? 

In the remainder of this study, we only consider stellar contribution to the intrinsic ionising photons budget, to be comparable with former studies, and we quantify the escaping  LyC contribution from the stars and the gas.

\section{Modelling LyC radiation transfer}
\label{section:data}
Throughout this work we use photoionisation models to estimate the escape of LyC nebular emission, which we now describe.

\subsection{\textsc{Cloudy} photoionisation models}
\label{subsection:models}
As a complement to the models presented in \cite{Inoue2010}, we run a grid of simple \textsc{Cloudy} photoionisation models to convergence, using BPASSv2.2.1 constant star formation stellar populations (with binaries) as intrinsic spectra. 
As a fiducial model, we choose to use a Salpeter IMF ($\alpha = -2.35$, with M$_{*,{\rm max}} = 100$ M$_\odot$), but we explore other choices in Sect.~\ref{section:LyC_dependence_on_intrinsic_spectra}  and Fig.~\ref{fig:intrinsic_LyC}, as well as in  Appendix~\ref{appendix}. We vary the ages and metallicities of the stellar SEDs, as summarised in Table~\ref{tab:stellar_SED_properties}, but most importantly for this work, we define the stopping criteria in \textsc{Cloudy} as the neutral hydrogen column density. We use the latter to explore the transition from LyC optically thin to optically thick media, and simulate density (\nhstop $= 10^{16}$ cm$^{-2}$, \fesc\ $\sim 100$\%) and  ionisation bounded (\nhstop $= 10^{19}$ cm$^{-2}$, \fesc\ $\sim 0$\%) \hii\ regions. A spherical cloud with chemical abundances and grains given by the \textsc{Cloudy} pre-defined ISM prescription \citep{Cowie1986,Savage1996} is assumed.

Once the photoionisation code has converged, \textsc{Cloudy} produces (among several other products) transmitted and net transmitted spectra. The transmitted spectra is the absorption spectra, which scales with the optical depth as 
\begin{equation}
\label{eq:transmitted}
L(\lambda) = L_0(\lambda)\times e^{-\tau(\lambda)}    
\end{equation}
where the transmitted radiation $L(\lambda)$ is given as a function of the incident radiation $L_0(\lambda)$ and the total optical depth of the medium $\tau(\lambda)$. The net transmitted spectra, on the other hand, also includes the diffuse emission that arises from the gas and dust particles within the nebula, which can also escape. Hereafter, we refer to the transmitted spectra as "stellar" and to the net transmitted spectra as "stellar+nebular".\\

Hereafter, we define the 5~Myr old \textsc{Cloudy} model with stellar metallicity $Z = 0.006$, and log(\nhstop\/)/[cm$^{-2}$] = 17.5 as fiducial. Unless otherwise specified, the ionisation parameter is fixed to -2.0, the density to 100 cm$^{-3}$, and the element abundances/grains are set to the predefined "ism" abundance in \textsc{Cloudy}.

\subsection{Wavelength dependence of the LyC nebular emission: Free-bound and line recombinations}
\label{subsection:LyCRT}

The radiation transfer of LyC photons through the nebula leads to two types of spectral features, illustrated in  Figure~\ref{fig:diffuse_effect}. The top panel of this figure shows the intrinsic (black curve), stellar (grey hatched area), and stellar+nebular (purple shaded area) LyC contributions emergent from our fiducial model. As can be seen in this plot, the stellar and stellar+nebular spectra can be significantly different, depending on wavelength. First, at wavelengths below the HeII, HeI and HI Lyman breaks wavelengths, $\lambda \sim 228, 504, 912$ \AA, the stellar+nebular spectrum is much stronger than the stellar spectrum (and in this particular example, even stronger than the intrinsic spectrum very close to the HI Lyman limit). This LyC nebular contribution, with its characteristic wing extending towards shorter wavelengths, is produced through free-bound recombinations, when helium and hydrogen nuclei catch fast-moving electrons directly to the ground state, leading to the emission of an ionising photon with energy corresponding to the ionisation energy plus the kinetic energy of the formerly free electron. This radiation is then produced by ionised media, and will escape if the residual opacity is low enough. Second, some recombination lines of helium have their wavelengths in the LyC range (e.g. 584.33\AA, 625.56\AA), boosting locally the number of hydrogen ionising photons, much above the intrinsic LyC stellar continuum. 

To illustrate the wavelength dependence of the LyC nebular contribution, the second panel shows, for the same fiducial model, the wavelength dependent escape fractions, i.e. the ratio between escaping (purple or grey curves in the top panel) and intrinsic (black curve) spectra of the top panel as a function of wavelength, for the stellar+nebular spectrum (in the pink shaded area) and the stellar spectrum (grey hatched area). The stellar+nebular escape fractions are equal to the stellar escape fractions except at wavelengths where free-bound or line recombinations arise. In these cases the stellar+nebular escape fractions are larger than the stellar ones, easily reaching values above 1 at specific wavelengths where the nebula produces a great amount of ionising photons. Along the same lines, the bottom panel of Figure~\ref{fig:diffuse_effect} shows the boost in flux due to the nebular contribution (ratio between the purple and grey curves of the top panel). The flux density of the stellar+nebular LyC spectrum is locally a few times to an order of magnitude brighter than the stellar LyC spectrum.

\section{Impact of LyC nebular emission}
\label{section:results}
In this section, we quantify the LyC nebular escape, and explore under which conditions LyC nebular emission is enhanced. For that, we consider the variation of escape fractions\footnote{Considering our former discussion on the several definitions of escape fractions in Section~\ref{section:fesc}, these are galactic escape fractions, directionally averaged given the spherically symmetric geometry, and averaged over different wavelengths ranges.}, and LyC luminosity increase, that we define as the \emph{boost factor}, the ratio between the stellar+nebular and the stellar-only luminosities. 

\begin{table}
        \centering
        \begin{tabular}{cc}
        \hline
        \noalign{\smallskip}
        Parameter & Values \\ 
        \noalign{\smallskip}
        \hline
        \noalign{\smallskip}
            Age [Myr] & 3, \textbf{5}, 10, 50\\ 
            $Z$ [BPASS] & 0.001, \textbf{0.006}, 0.014, 0.030\\
            log(\nhstop)/[cm$^{-2}$]& $16, 16.5, 17, \textbf{17.5}, 18, 18.5, 19$\\
        \noalign{\smallskip}
        \hline
        \noalign{\smallskip}
        \end{tabular}
        \caption{General properties of the intrinsic spectra used in the models and the \nhstop\ values adopted. All are BPASS stellar populations with binaries and a constant star formation law is assumed. The parameters corresponding to the fiducial model are bold-faced. \textsl{Row 1:} age of the population. \textsl{Row 2:} metallicity of the stellar population, where $Z = 0.020$ corresponds to solar metallicity. \textsl{Row 3:} logarithm of the neutral hydrogen column values used as stopping criteria in \textsc{Cloudy}.}
        \label{tab:stellar_SED_properties}
    \end{table}

\subsection{Dependence on the physical conditions in the cloud}
In this subsection, we consider a sample of intrinsic stellar continua as the source of the LyC radiation, and we vary the conditions in the cloud: first the column density which has the strongest impact; then the density and ionisation parameter of the cloud.

\subsubsection{Impact of the column density of the cloud}
  
Figure~\ref{fig:Lybump} shows the variation with \nhstop\ of three usual definitions of galactic escape fractions\footnote{Given our spherically symmetric setup, these galactic escape fractions can be considered as directionally averaged, or estimated along an arbitrary line of sight, since they are equal for our problem.}: the integrated \fesc\ (in the range $\lambda = 0 - 912$ \AA\/), the \fesc\ calculated at $\lambda = 800 - 912$ \AA\/, and the monochromatic \fesc\ at 900 \AA, with (purple) and without (grey) the inclusion of the nebular emission, for BPASS models spanning a range of ages and metallicities described in Table~\ref{tab:stellar_SED_properties}. 

\begin{figure}
    \centering
    \includegraphics[width=1\columnwidth]{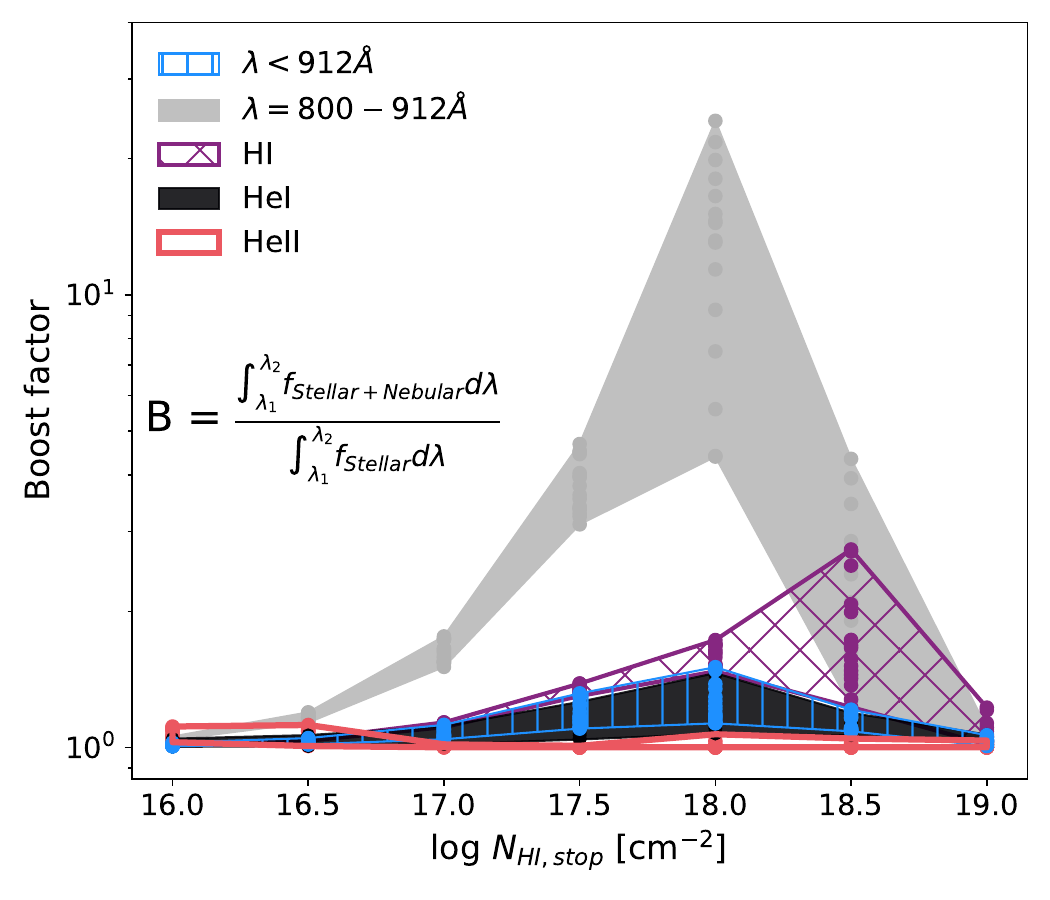}
    \caption{Strength of the LyC boost as a function of \nhstop\/, defined as the integral of the stellar+nebular flux density over the integral of the stellar flux density in a given wavelength range. The integration limits are $\lambda < 912$ \AA\/ (blue hatched area), $\lambda = 800-912$ \AA\/ (grey shaded area),  $\lambda = 504-912$ \AA\/ (purple hatched area; "HI"), $\lambda = 228-504$ \AA\/ (black area; "HeI"), and $\lambda < 228$ \AA\/ (red area; "HeII"), respectively. The regions correspond to the hydrogen and helium ionisation potentials, in addition to the spectral region close to the Lyman limit. There is a clear increase in the boost factor for log \nhstop\//[cm$^{-2}$] = 18.0, around 900 \AA\/ (we note that the small denominator plays a part in this increase). As expected, the largest difference is found for the bin close to the Lyman limit ($\lambda = 800-912$ \AA\/). The vertical spread is driven by age and stellar metallicity.}
    \label{fig:boost}
\end{figure}

First, we notice that nebular emission is important only at intermediate opacities, around $17 <$ log(\nhstop/cm$^{-2}$) $< 18$ . When the ISM is completely transparent to LyC, log(\nhstop/cm$^{-2}$) $\leq 16$, then everything escapes: since there is no significant LyC absorption in the ISM, there is no radiation transfer, and the escaping spectrum is the same as the intrinsic spectrum. So the grey and purple curves converge to each other at log(\nhstop/cm$^{-2}$)=16 for both the monochromatic and the wavelength integrated escape fractions. Second, at high opacity, all LyC photons are absorbed: they are reshuffled in wavelengths, but then re-absorbed again, as long as they are still ionising, and finally all LyC photons are converted into non-ionising nebular recombination lines, that is why the purple and grey curves converge to each other also at high N$_{\rm{HI}}$ (log(\nhstop/cm$^{-2}$) $= 19$). At intermediate opacities, however, a significant fraction of the intrinsic LyC spectrum, that is re-emitted in the LyC range through free-bound or line recombination, will eventually escape the medium. 
While it only slightly affects the integrated escape fractions, it produces significant differences when measured at $900$ \AA, as detailed below.

The left panel shows that the grey curves showing escape fractions estimated as \fesc\/ = exp($-\tau$) are always below the purple curves, which take into account the escape of nebular LyC photons. Neglecting the escape of nebular LyC continuum, as done so far in cosmological simulations of galaxy formation, underestimates the wavelength-integrated escape fractions, with a maximum effect up to a factor of 1.5 for leaking star-forming regions in front of an ISM of log(\nhstop/cm$^{-2}$) = 17.5.

As can be seen in the middle panel, this effect is strongly amplified at $\lambda \sim 800 - 912$ \AA\/, where most LyC detections have been obtained so far, due to the nebular HI free-bound recombinations. The LyC nebular contribution is particularly important at column densities of \nhstop\ $\sim 10^{17} - 10^{18}$ cm$^{-2}$, where escape fractions, calculated as the ratio between the intrinsic stellar spectrum and the emergent LyC stellar+nebular spectrum over the same wavelength range, are increased by more than a factor 3, with respect to escape fractions of the transmitted stellar spectrum. 

Finally, as shown in the right panel for the fiducial model, the monochromatic escape fractions at 900 \AA\/  reach values above 1: there are more ionising photons escaping at these wavelengths than produced intrinsically by the stars, for intermediate opacities in the cloud (\nhstop\ $\sim 10^{17} - 10^{18}$ cm$^{-2}$). 
So,  
the spectral region around 900 \AA\/ appears like the worst wavelength range to estimate the total LyC photons budget escaping that galaxy, since it implies taking into account the nebular emission contribution, which depends on the physical conditions in the cloud, as explored further below. We propose to instead observe the LyC emission at shorter wavelengths, ideally $630< \lambda < 800$ \AA\ (see Section ~\ref{section:obs}), i.e. far from any LyC radiation transfer effects, as illustrated on Figure~\ref{fig:diffuse_effect}.

\begin{figure}
    \centering
    \includegraphics[width=1\columnwidth,trim={0 0 0 0}]{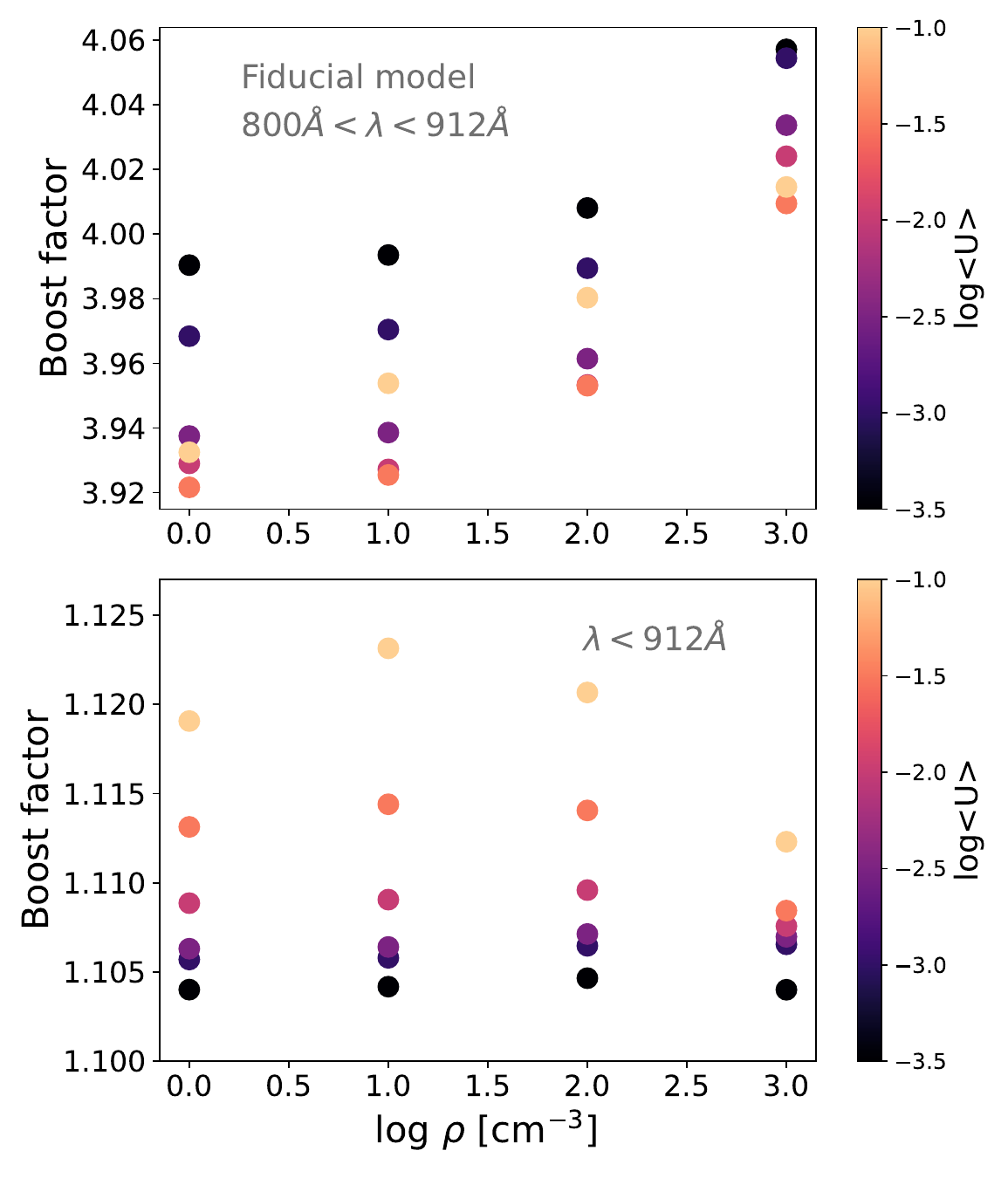}
    \caption{Variations of the LyC nebular emission with the cloud density, colour-coded by ionisation parameter, for the fiducial model. \textsl{Top panel:} boost integrated at wavelengths between 800 and 912 \AA\/. \textsl{Bottom panel:} boost integrated at all wavelengths below 912 \AA\/. The strength of the boosts show only a slight dependence with density and ionisation parameter of the nebula, for a fixed stellar model and \nhstop\/.} 
\label{fig:physcond_LyC}
\end{figure}

\begin{figure}
    \centering
    \includegraphics[width=1\columnwidth,trim={0 0 0 0}]{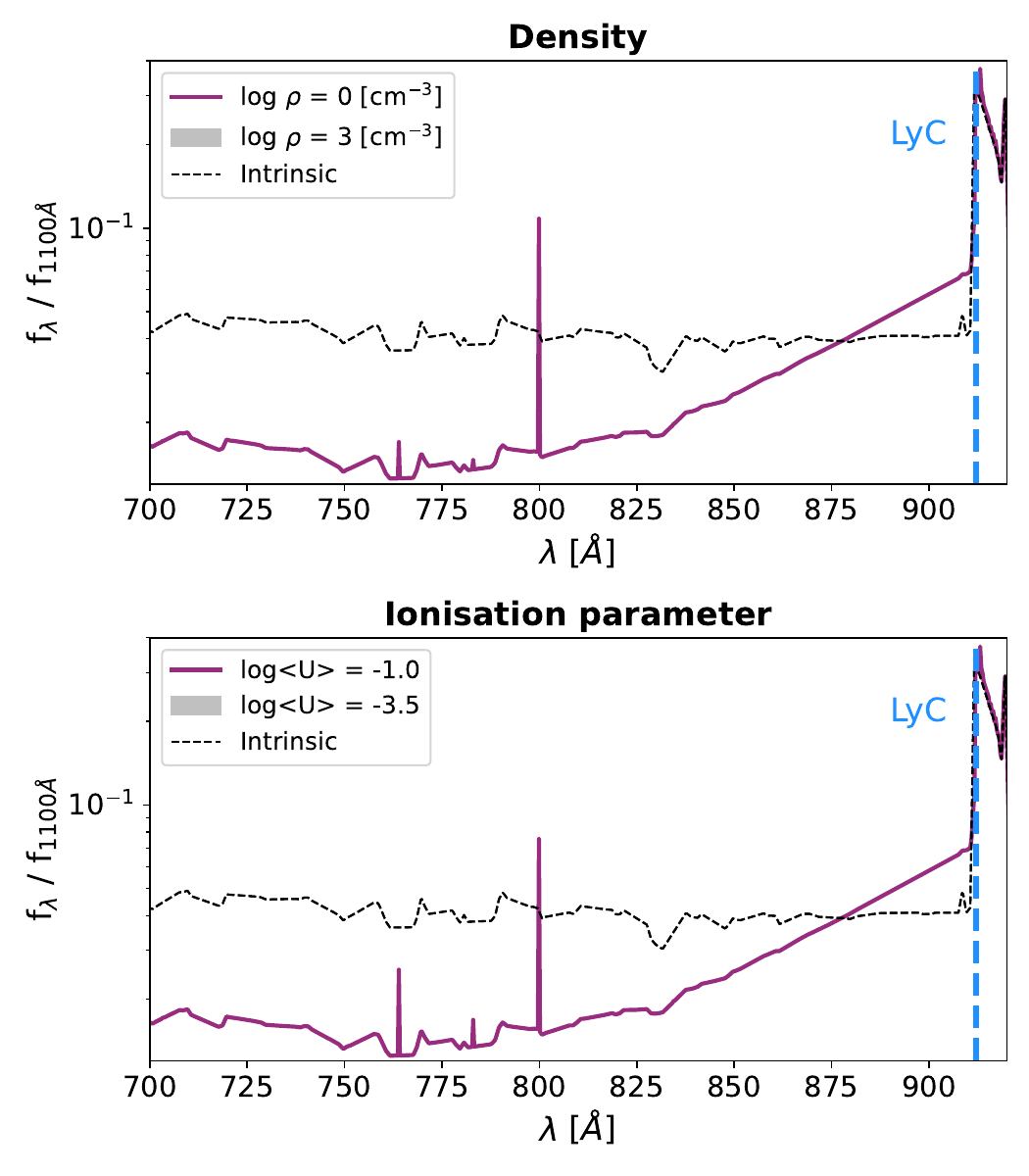}
    \caption{Effect of nebular parameters on the shape of stellar+nebular emission. The dashed blue vertical line marks the Lyman limit. \textsl{Top panel:} effect of changing the density of the nebula, adopting the fiducial model and a fixed log<U> of -2.5. \textsl{Bottom panel:} effect of changing the ionisation parameter in the nebula, for the fiducial model and a density of 100 cm$^{-3}$. The dependence of the shape of the ionising spectra on density and ionisation parameter is small.}  
    \label{fig:parameters_bump_effect}
\end{figure}

Escape fractions are a primary scope of research, but as demonstrated in Figure~\ref{fig:Lybump}, they are extremely dependant on the exact wavelength range over which they are estimated, and not directly proportional to the total LyC photons budget. A complementary method to quantify the importance of LyC nebular emission is to calculate how the boost in LyC luminosity varies with \nhstop\/. Similarly to escape fractions, we consider the total boost, integrated over all ionising wavelengths:
\begin{equation}
\text{Boost} = \frac{\int_0^{912} F_{\lambda,\rm stellar+nebular}d\lambda}{\int_0^{912} F_{\lambda, \rm stellar}d\lambda}.
\label{eq:boost}
\end{equation}
In Figure~\ref{fig:boost} we present the variations with \nhstop\ of this total boost in blue, and boosts over convenient wavelength bins: between $\lambda = 800-912$ \AA\/ (grey shaded area), and at wavelengths corresponding to mostly HI, HeI and HeII ionising photons (red hatched, black filled and purple regions, respectively). Again, it is only at intermediate opacities that this effect is important. The wavelength-integrated boost can be up to a factor of 1.5 at log(\nhstop\/)/[cm$^{-2}$] = 18. It is evident that the boost is most important close to the Lyman limit. On the observed wavelength range, it varies from 4 to 10 at log(\nhstop\/)/[cm$^{-2}$] = 18, and is around 3 at log(\nhstop\/)/[cm$^{-2}$] = 17.5, which implies that 
a large amount of LyC photons in the detections reported so far in the literature might have a nebular origin, rather than stellar.Conversely, the luminosity boost in the He~II bin is negligible. We discuss implications for simulations and observations in Sections~\ref{section:sim} and \ref{section:obs}. 

\begin{figure*}
    \centering
    \includegraphics[width=2\columnwidth]{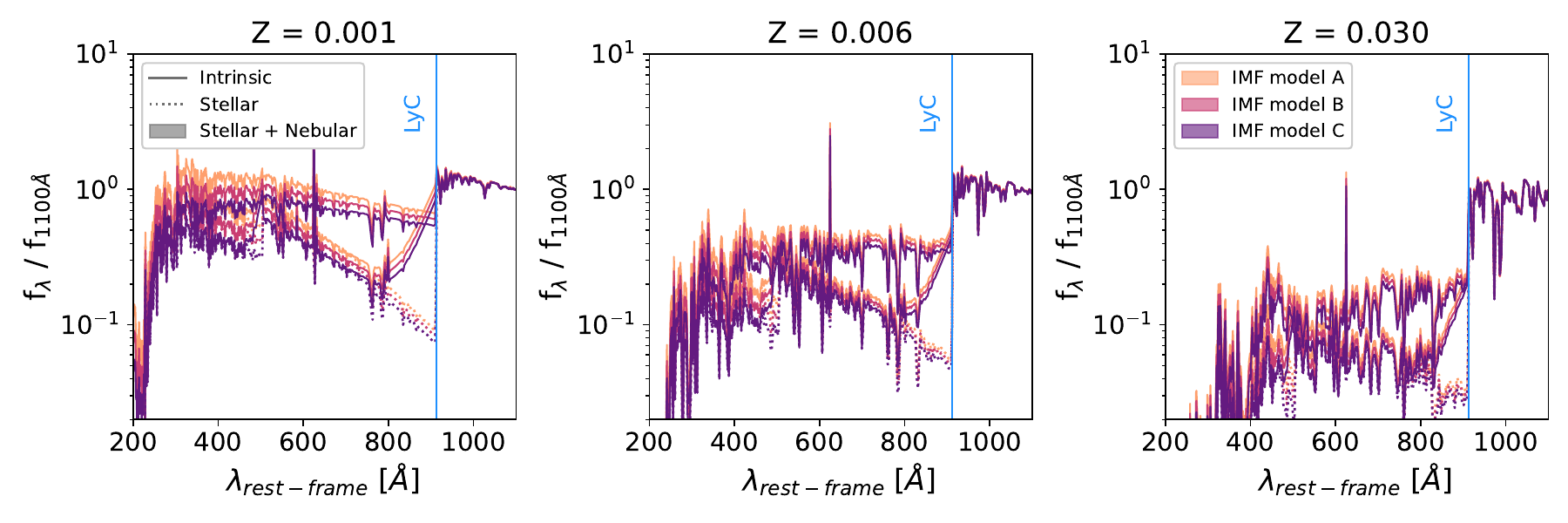}
    \caption{Shape of LyC emission depending on initial mass function (IMF) and metallicity, for a stellar population of age $\sim$5 Myr formed in a burst. Each panel shows the intrinsic (filled lines), stellar (dotted lines) and stellar+nebular (shaded area) flux densities, assuming a \textsc{Cloudy} model with log(\nhstop\/)/[cm$^{-2}$] = 17.5, for a fixed metallicity as indicated in the title, and colour-coded by the IMF model. All values have been normalised to the flux density at $\lambda = 1100$ \AA\/. For illustrative purposes, we choose to show IMFs with an upper mass-cut of 300 M$_{\odot}$, but note that lowering this quantity to 100 M$_{\odot}$ does not produce a noticeable difference in the shape of the spectra. \textsl{Model A:} $\alpha_2 = -2.00$. \textsl{Model B:} $\alpha_2 = -2.35$. \textsl{Model C:}  $\alpha_2 = -2.70$. Where $\alpha_2$ is the higher mass (0.5 < M < 300 M$_{\odot}$) exponent, assuming a broken power law IMF (the lower mass exponent, $\alpha_1$ is fixed at $-1.30$ and has no effect on the resulting UV spectra). It can be seen that for a fixed age and metallicity, the choice of IMF results in only a slight difference in the shape of the ionising spectra, but that the nebular emission produces a flux excess at $\lambda = 800 - 900$ \AA\/ in every case. }
    \label{fig:intrinsic_LyC}
\end{figure*}

From now on, we fix \nhstop\/ to log(\nhstop)/[cm$^{-2}$] = 17.5, and we explore secondary effects by varying the physical conditions in the cloud, and the intrinsic LyC stellar spectrum. 

\subsubsection{Dependence on the physical conditions in the cloud}
Here we explore how the conditions of the cloud impact
the resulting nebular boost of ionising photons, defined in Equation~\ref{eq:boost}. For this purpose, we create a grid of \textsc{Cloudy} models based on the fiducial model, but varying the total hydrogen density and ionisation parameter of the cloud (log($\rho$/cm$^{-3}$) between 0 and 3, log<U> from -3.5 to -1.0). Figure~\ref{fig:physcond_LyC} shows how the boost factor depends on these properties: in the top panel we show the boost at wavelengths between 800 and 912 \AA\/, while in the bottom panel we show the total boost (integrated at all wavelengths below 912 \AA\/). We find that the boost factor in the region close to the Lyman limit ($800$ \AA\ $< \lambda < 912$ \AA\/) has a dependence on density, while the total boost ($\lambda < 912$ \AA\/) shows a stronger dependence on ionisation parameter. The former is due to the increase of collisions in the cloud, which allows for more recombination and thus nebular emission. The latter is a natural consequence of the definition of log<U>, which in \textsc{Cloudy} represents the ratio between the hydrogen-ionising photons and the total hydrogen density. We note that the dependence in both cases is weak, resulting only in variations of the order of 1\%. We thus conclude that the strengths of the boosts, when \nhstop\/ is fixed, relate primarily to the intrinsic stellar model assumed, and only secondarily on the conditions of the cloud such as density and ionisation parameter.

Whereas Figure~\ref{fig:physcond_LyC} shows the effect of these properties on the boost factors, in Figure~\ref{fig:parameters_bump_effect} we explore how the shape of LyC emission changes  due to them, specifically at wavelengths near the Lyman limit. We use the same \textsc{Cloudy} grid described above and show cases with fixed log<U> and varying density (top panel), and fixed density but variable log<U> (bottom panel). As in the case of the boost factors, the shape of the ionising stellar+nebular spectra has but a slight dependence on these parameters. However, we direct the readers to the models presented in \cite{Inoue2010}, where the shape of the LyC emission takes on a more peaked shape as temperature decreases.

\subsection{Dependence on the intrinsic stellar spectra}
\label{section:LyC_dependence_on_intrinsic_spectra}
The LyC spectrum of stars is not observationally constrained. Only two massive stars have been observed in their LyC wavelength range \citep{Craig1997,Erickson2021}, and there are several uncertainties concerning the global shape, hardness, strength, and time evolution of the ionising radiation emitted from stars. The LyC photons budget from a single stellar population is predicted to depend on the age of the population, linked to the star formation history, the initial mass function and initial mass cut-off \citep{Leitherer1999}, the  metallicity of the stars \citep{Schaerer2003}, or the binary fraction \citep{Eldridge2017}. We explore in this section the impact of these assumptions, or uncertainties, on the LyC nebular emission, and wavelength integrated LyC escape fraction and boost.

As can be seen in Figure~\ref{fig:intrinsic_LyC}, for a fixed age and metallicity, the choice of IMF results in only a slight difference in the shape of the ionising spectra. Therefore, having a negligible effect on both the total escape fractions and boost factor.  
We note that this conclusion was reached by using the IMFs available in BPASS (with a range in upper mass slope of $\alpha_2 = -2.7$ to $-2.0$, where $-2.35$ corresponds to the Salpeter IMF), and that this might change if a more extreme top heavy IMF were adopted instead. An exploration of other upper mass slopes for Starburst99 \citep[SB99; ][]{Leitherer1999} models formed in a burst can be found in Appendix~\ref{appendix}. 
   
\section{Discussion}
\label{section:discussion}

In this section, we discuss the implications of our results for observations and simulations; and the limits of our approach to estimate LyC nebular contributions from idealised \textsc{Cloudy} runs. Finally, we mention possible other sources of ionising radiation in galaxies.

\subsection{Impact of LyC nebular emission in simulations}
\label{section:sim}
\begin{figure*}
   \centering
    \includegraphics[width=0.5\textwidth,trim={0 0 0 0}]{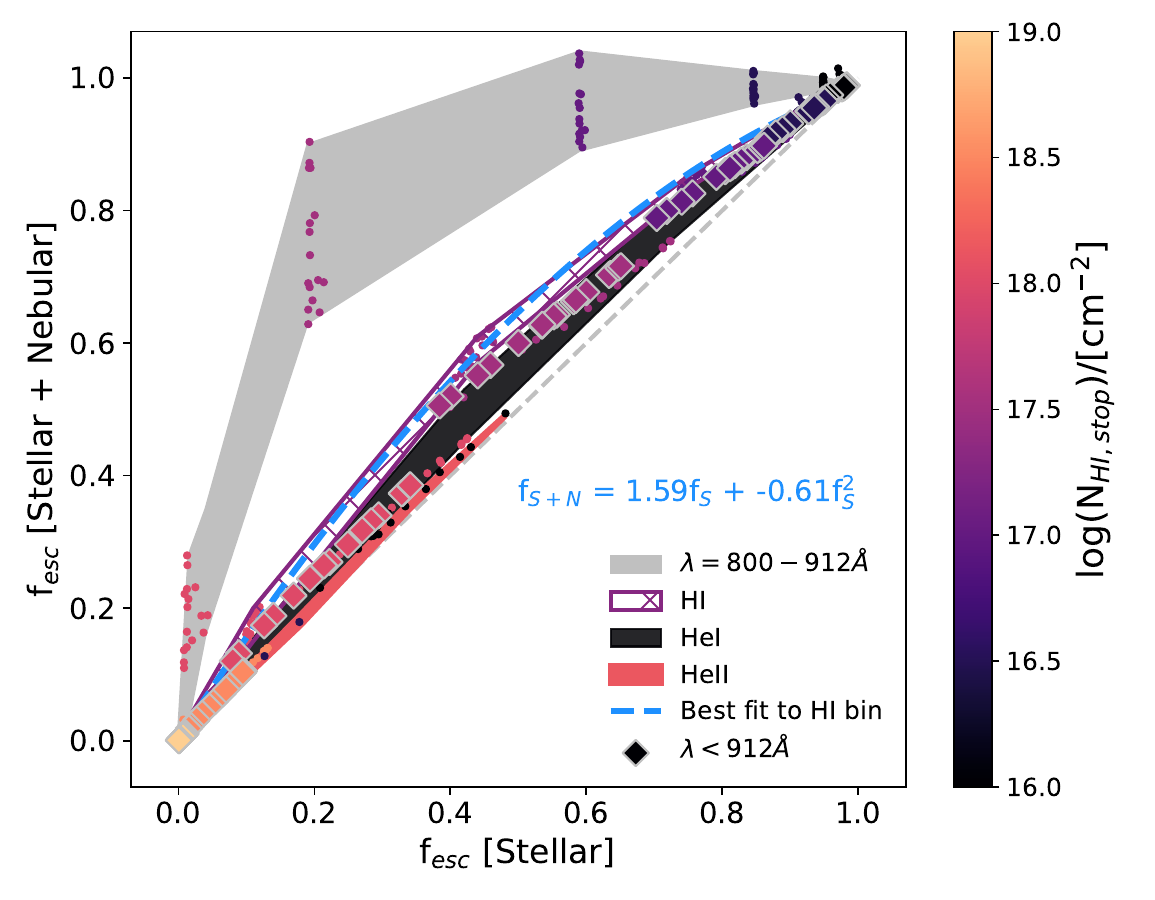}
    \includegraphics[width=0.45\textwidth,trim={0 0 0 0}]{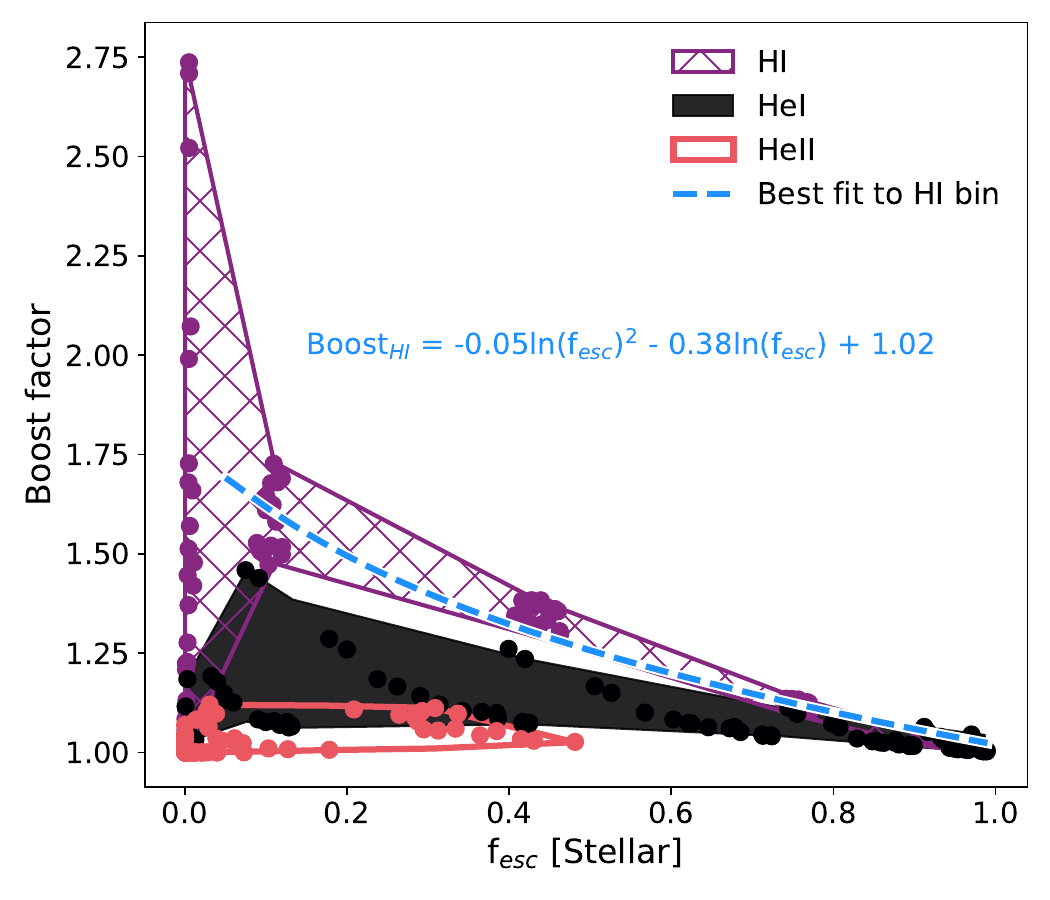}
    \caption{Strength of the nebular LyC emission versus stellar escape fractions, 
    the latter corresponds to the escape fraction usually computed in simulations, for the three LyC bins usually considered: HI-only ionising photons (purple, $504 < \lambda < 912$ \AA), HeI and HI ionising photons (black, $228 < \lambda < 504$ \AA), HeII, HeI and HI ionising photons (red, $\lambda < 228$ \AA). \textsl{Left panel:} escape fractions per wavelength bin as a function of the stellar escape fraction. The bins are shown as shaded areas with points colour-coded by \nhstop\/, and the escape fractions at all wavelengths below 912 \AA\/ are shown as diamonds with grey edges. The best fit to the HI bin is shown as a blue dashed line, where f$_{\rm{S+N}}$ and f$_{\rm{S}}$ denote stellar+nebular, and stellar escape fractions, respectively.
    \textsl{Right panel:} boost factor per wavelength bin as a function of the stellar escape fraction. The best fit relation is shown as a blue dashed line.}
              \label{fig:boostPerBin}
   \end{figure*}

It is often assumed that stars are the primary sources of ionising radiation in galaxies. When calculating escape fractions from galaxies, so far cosmological simulations of reionisation usually compute the \emph{transmitted} LyC escape fractions as:
\begin{equation}
f_{\rm esc, stellar} = \frac{\int_0^{912} F_{\lambda_0}\times e^{-\tau(\lambda)}d\lambda}{\int_0^{912} F_{\lambda_0}d\lambda},
\end{equation}
which we call 'stellar' in this study, neglecting LyC radiation transfer in the ISM, and the possible escape of additional LyC nebular emission. We propose in this section possible correction factors to estimate stellar+nebular galactic escape fractions from the stellar values.

Figure~\ref{fig:boostPerBin} shows the evolution of escape fractions (left panel) and boost factors (right panel) with respect to the stellar escape fraction, sliced over three wavelength bins that are often considered in cosmological radiation hydrodynamics simulations  (such as e.g. THESAN and SPHINX): the first bin containing hydrogen-only ionising photons (HI in purple, $504 < \lambda < 912$ \AA); the second bin containing photons that can ionise both hydrogen and helium (HeI in black, $228 < \lambda < 504$ \AA), and the last bin that can ionise hydrogen, helium, and He$^{+}$ (HeII in red, $\lambda < 228$ \AA). The HeI and HeII escape fractions and boosts are marginally affected by LyC nebular emission, only the first bin (HI) requires a correction. 
The luminosity boost is inversely proportional to the stellar escape fraction, but still reaches 1.5 for stellar escape fractions of 10\% from simulated galaxies. We propose formulae to correct the (escape fraction and) luminosity boost of the HI bin, when the stellar escape fraction has been estimated. 

These corrections may help estimate the contribution of nebular LyC escape to the total ionising photons budget from galaxies, i.e. to the cosmic escape fraction. However,  in order to investigate the impact of these photons on the reionisation history of the Universe, full RHD simulations need to be re-run.  

\subsection{Impact of LyC nebular emission in observations}
\label{section:obs}

   \begin{figure}
   \centering
   \includegraphics[width=1\columnwidth]{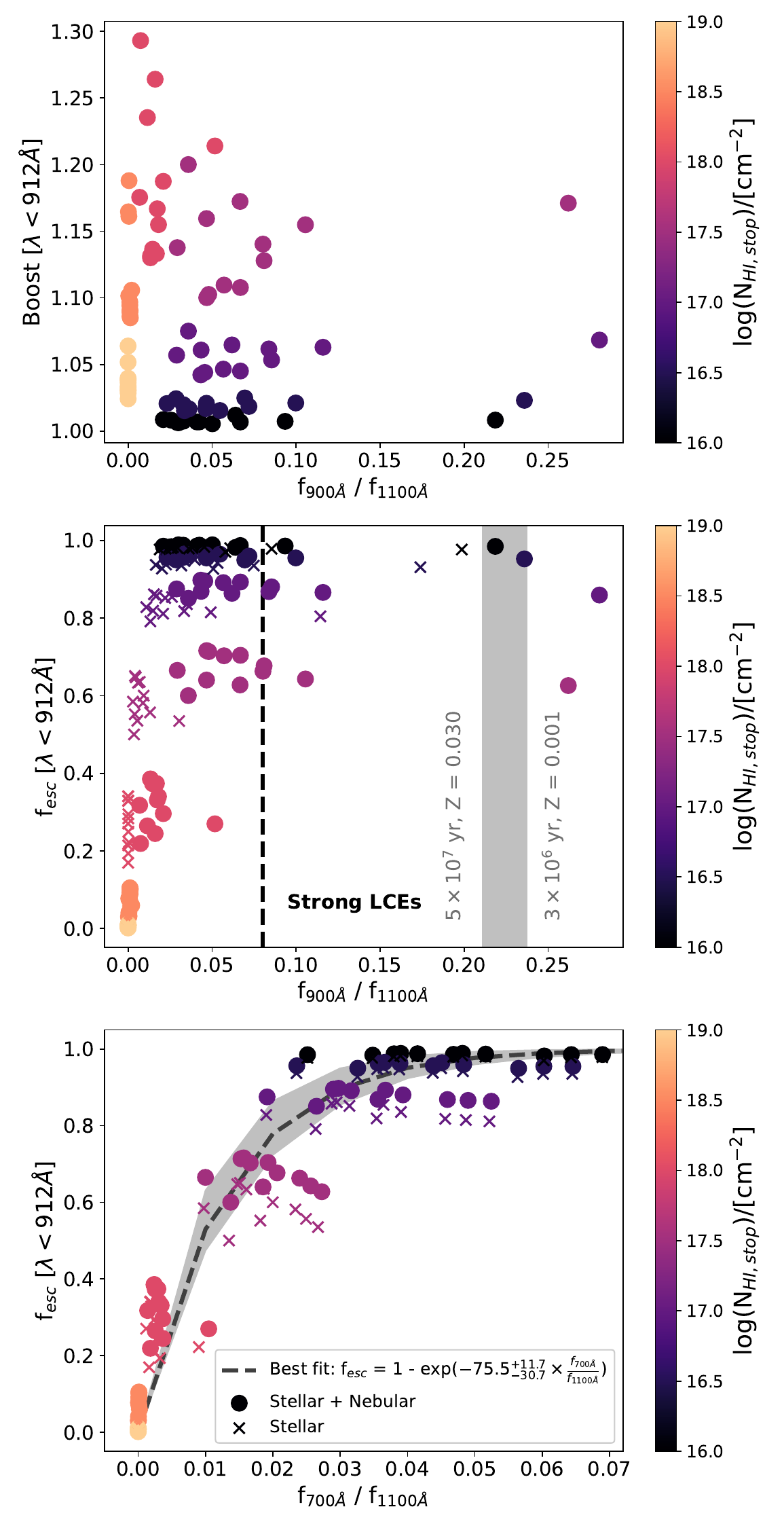}
   \caption{Strength of the LyC nebular emission versus directly observable ionising over non-ionising flux ratios. \textsl{Top panel:} Boost factor (integrated at all wavelengths below 912 \AA\/)  as a function of the ratio between flux densities at  $\lambda = 900$ and $1100$ \AA\/, colour-coded by log(\nhstop\/). \textsl{Middle panel:} \fesc\ versus ratio between flux densities at $\lambda = 900$ and $1100$ \AA\/, colour-coded by log(\nhstop\/). Strong LyC emitters from the LzLCS \citep[$> 5\sigma$ LyC detection and \fesc\/ $> 5$\%;][]{Flury2022b} lie to the right of the dashed black line. The spread in flux ratio for a fixed \nhstop\ is due to the age and metallicity of the incident spectra, the former being the primary driver of the dispersion. The models that include nebular emission (stellar+nebular, circles), show a significant increase in the flux ratio in the range \nhstop\ $\sim 10^{17} - 10^{18}$ cm$^{-2}$, in comparison to the stellar (transmitted, crosses) ratios. \textsl{Bottom panel:} same as middle panel but with the ratio between flux densities at $\lambda = 700$ and $1100$ \AA\ instead. At $\lambda \sim 700$ \AA\ the nebular contribution is minimal, therefore, the ratios considering only stellar and stellar+nebular emission agree. The best fit to the data is given as a grey dashed curve and shaded region. The errors were determined by bootstrapping 1000 iterations taking two thirds of the sample in each iteration, randomly selected.} 
              \label{fig:4aFlury2022b}
   \end{figure}

Since 2016, LyC detections have been reported in the literature, both at low redshift \citep[89 observations, among which 55 detections, at $z\sim0.3$,][]{Izotov2016a,Izotov2016b, Izotov2018a, Izotov2018b, Izotov2021, Wang2019, Flury2022a} and at intermediate redshifts \citep[a few dozen of galaxies at $z\sim3-4$]{Vanzella2016, Steidel2018, Marques-Chaves2021, Marques-Chaves2022, Rivera-Thorsen2019,Kerutt2023}. 
The escape fractions derived from these detections are  galactic, line-of-sight estimates, over a restricted wavelength range, depending on each observation. 
Most of the observations probe LyC around rest-frame $\lambda = 800 - 900$ \AA, with some exceptions. For example, LyC was observed in AUDFs01 at rest-frame $\lambda \sim 600$ \AA\/ \citep{Saha2020}. As discussed in Sect.~\ref{section:results}, the wavelength range $\lambda = 800 - 900$ \AA\/ is the most impacted by LyC nebular emission. We investigate in this section how to recover the cosmologically relevant wavelength integrated escape fraction, directly from observables. 

The ionising over non-ionising UV radiation, $\text{f}_{\lambda 900} / \text{f}_{\lambda1100}$, is a direct observable, but it is impacted by the strength of the free-bound nebular emission. As shown on the top panel of Figure~\ref{fig:4aFlury2022b}, for a given observed flux ratio between 0.01 and 0.1, the nebular LyC contribution varies from 0 to 30\% of the total LyC luminosity escaping a galaxy, depending on \nhstop. 
On the middle panel we show that, for a given value of \fesc\/ > 60\%, the observed ratio $\text{f}_{\lambda 900} / \text{f}_{\lambda1100}$ can take any value between 0 and 0.3, driven by stellar age and metallicity. If BPASS LyC stellar templates are taken at face values, $\text{f}_{\lambda 900} / \text{f}_{\lambda1100}$ ratios as low as 0.03 can still trace escape fractions of 100\%. In extreme cases, the observable ratio $\text{f}_{\lambda 900} / \text{f}_{\lambda1100}$ reaches values above the range allowed by intrinsic stellar (grey area) $\text{f}_{\lambda 900} / \text{f}_{\lambda1100}$ ratios from the BPASS library ($\lesssim 0.25$). 
This indicates that the ratio $\text{f}_{\lambda 900} / \text{f}_{\lambda1100}$ cannot be used to estimate the global escape fraction of ionising radiation from galaxies. 

To circumvent this issue, we propose to measure LyC fluxes around $\lambda = 700$ \AA\/. The bottom panel of Figure~\ref{fig:4aFlury2022b} shows the total escape fraction but as a function of $\text{f}_{700\lambda} / \text{f}_{1100\lambda}$ instead. In this case the stellar and the stellar+nebular agree and can be used to estimate the global escape fractions. We find a best-fit relation to all the data points given by:
\begin{equation}
    \text{f}_{\rm{esc}} = 1 - \exp\bigg(-75.5^{+11.7}_{-30.7}\times \frac{\text{f}_{700\text{\AA\/}}}{\text{f}_{1100\text{\AA\/}}}\bigg).
\end{equation}
Therefore, by avoiding the spectral region contaminated by emission lines and free-bound nebular emission, it is possible to estimate the total escape fraction.

\subsection{Limitations of our approach}
Our individual \textsc{Cloudy} models are idealised spherically symmetrical configurations, that may be seen as representing single star forming regions, while galaxies contain thousands of star forming clumps, with different ages and metallicities, surrounded by ISM gas spanning a broad range of column densities. 
In principle, galaxies may be better modelled by linear combinations of individual \textsc{Cloudy} models, except if the UV spectra of LyC leakers are dominated by a single leaking star-forming region that overshines the rest of the galaxy. For example, a more complex approach to photoionisation modelling is presented in \citep{Marconi2024}, where weighted combinations of single-cloud photoionisation models are used to determine accurate galaxy properties. Some LyC leakers indeed have rather compact UV morphologies \citep{Izotov2021}, which is particularly interesting since observed escape fractions seem to correlate with compactness, or surface star-formation rates \citep{Flury2022b}. 

In any case, this study should only be considered as a first step to enlight the potential importance of LyC nebular emission. The next step would be to simulate the LyC emissivity of the ISM gas in virtual galaxies, and process the LyC radiation transfer of these photons on top of the stellar LyC radiation to estimate more accurate galactic escape fractions. The ultimate goal being to estimate the cosmic escape fraction, allowing to assess the real contribution of LyC nebular emission to the reionisation of the Universe. 

\subsection{Other sources of ionising radiation in galaxies}

As already mentioned, AGNs are usually considered as the second type of possible contributors to cosmic reionisation \citep{Madau2015}. However, high redshift observations seem to indicate that quasars are too rare at high redshift to be the main sources \citep{Onoue2017},
and simulations have also shown that their contribution to the cosmic UV background becomes important at z$< 4$ \citep{Trebitsch2021}. 

Taking into account interactions from binary stars, which allows to emit more ionising photons over longer timescales, has proven a key ingredient to reionise the SPHINX Universe \citep{Rosdahl2018}, and the BPASS stellar library \citep{Eldridge2017} is now used as the default SED library in state-of-the-art RHD simulations of Reionisation, motivating us to choose these stellar models for our study. X-ray binaries also have a very energetic spectrum \citep{Simmonds2021}, and high mass X-ray binaries could have played a role in ionising the early Universe \citep{Mirabel2011, Eide2020}, however, they are not yet included in SED models. 

In addition to these compact energetic sources, the diffuse ISM gas itself could provide supplementary ionising photons. (1) Through photoionisation from cosmic rays, although they arise in the densest parts of the ISM, from which it is unlikely that they escape.   
(2) The ISM gas can be collisionally photoionised by shocks due to turbulent motions, and a fraction of these photons could escape galaxies. In \cite{Johnson2011}, they estimate the contribution of shocks driven by Super Novae explosions in distant galaxies to cosmic reionisation, and in \cite{Dopita2011} they propose cosmological accretion flows as the drivers of collisional photoionisation of the circum-galactic medium, and since the productions sites of these photons would be in the outskirts of galaxies, a large fraction of this ionising radiation may escape. But these contributions are usually considered as subdominant. 

As a summary, in addition to the LyC nebular emission discussed in this study, AGNS, X-ray binaries and shocks all exist in galaxies, and they all produce ionising radiation. The potential contribution of these processes may not be dominant, but could provide the necessary supplement of photons to ease the reionisation history of the Universe.

\section{Conclusions}
\label{section:conclusions}

In this work, we studied the impact of nebular LyC emission on estimations of the escape fractions of ionising photons from galaxies, and their corresponding boosts in LyC luminosities, by running \textsc{Cloudy} models for ionising sources and cloud conditions spanning a range of properties, and varying the hydrogen column density \nhstop\/. Our main results are the following:
\begin{itemize}
    \item Due to (free-bound or lines) recombination processes, the ionised gas in galaxies emits LyC radiation, with a complex spectral distribution. This leads to strong and rapid variations of LyC luminosities and escape fractions over the LyC wavelength range, so that their estimates are highly wavelength dependent.
    \item The amount and the spectral distribution of the free-bound nebular emission depends on the incident LyC spectral shape, 
    and strongly on the column density of gas in front of the ionising sources, making it difficult to predict and to take into account in observations of LyC emission from galaxies taken at $\sim 900$ \AA\/. 
    \item On the other hand, the LyC emission at $\sim 700$ \AA\/ is devoid of nebular emission, and we propose an equation to relate the observable f$_{\lambda 700}$/f$_{\lambda 1100}$ flux ratio to  the cosmologically relevant wavelength-integrated LyC escape fraction in equation 5.
    \item Cosmological simulations of the Epoch of Reionisation usually  neglect the emission and escape of LyC photons from the gas in galaxies. From our numerical experiments, the luminosities of the HI photon packet is typically boosted by a factor 2 and the HeI photon packet by up to 50\%, these boosts being inversely proportional to the stellar escape fractions.
\end{itemize}

Thanks to these very idealised simulations, we have shown that LyC nebular emission may significantly increase the LyC luminosities and escape fractions of galaxies. In order to assess the role of these nebular LyC photons in the reionisation process,
the next steps will be to run cosmological simulations that take into account this production channel of ionising radiation. 

Observations of galaxies at $\sim 700$ \AA\/ rest-frame are very challenging for the samples of known Lyman continuum Leakers. This spectral range may be accessible with HST/COS for leaking galaxies at $z \sim 1$, that still need to be identified.  The modelling of actual observations at $900$ \AA\/ may allow to directly probe the spectral dependence of LyC to nebular emission, and infer the physical conditions in the ionised gas from the extend of the free-bound emission wing. Future instruments like HWO will open further avenues in our exploration of the LyC spectral distribution of galaxies.

\section*{Acknowledgements}
We acknowledge the numerous and inspiring discussions with S. de Barros, and his first  simulations on this topic. 
C.S., A.V. and T.G acknowledge support from SNF Professorship PP00P2\_176808 and its prolongation PP00P2\_211023. C.S. also acknowledges support by the Science and Technology Facilities Council (STFC) and by the ERC through Advanced Grant number 695671 ‘QUENCH’, and by the UKRI Frontier Research grant RISEandFALL.
A.K.I. acknowledges support from JSPS KAKENHI Grant Number JP23H00131.

\section*{Data Availability}
The data underlying this article will be shared on reasonable request to the corresponding author.



\bibliographystyle{mnras}
\bibliography{bib} 

\begin{thebibliography}{}
\makeatletter
\relax
\def\mn@urlcharsother{\let\do\@makeother \do\$\do\&\do\#\do\^\do\_\do\%\do\~}
\def\mn@doi{\begingroup\mn@urlcharsother \@ifnextchar [ {\mn@doi@} {\mn@doi@[]}}
\def\mn@doi@[#1]#2{\def\@tempa{#1}\ifx\@tempa\@empty \href {http://dx.doi.org/#2} {doi:#2}\else \href {http://dx.doi.org/#2} {#1}\fi \endgroup}
\def\mn@eprint#1#2{\mn@eprint@#1:#2::\@nil}
\def\mn@eprint@arXiv#1{\href {http://arxiv.org/abs/#1} {{\tt arXiv:#1}}}
\def\mn@eprint@dblp#1{\href {http://dblp.uni-trier.de/rec/bibtex/#1.xml} {dblp:#1}}
\def\mn@eprint@#1:#2:#3:#4\@nil{\def\@tempa {#1}\def\@tempb {#2}\def\@tempc {#3}\ifx \@tempc \@empty \let \@tempc \@tempb \let \@tempb \@tempa \fi \ifx \@tempb \@empty \def\@tempb {arXiv}\fi \@ifundefined {mn@eprint@\@tempb}{\@tempb:\@tempc}{\expandafter \expandafter \csname mn@eprint@\@tempb\endcsname \expandafter{\@tempc}}}

\bibitem[\protect\citeauthoryear{{Atek} et~al.,}{{Atek} et~al.}{2014}]{Atek2014}
{Atek} H.,  et~al., 2014, \mn@doi [\apj] {10.1088/0004-637X/786/1/60}, \href {https://ui.adsabs.harvard.edu/abs/2014ApJ...786...60A} {786, 60}

\bibitem[\protect\citeauthoryear{{Atek}, {Richard}, {Kneib}  \& {Schaerer}}{{Atek} et~al.}{2018}]{Atek2018}
{Atek} H.,  {Richard} J.,  {Kneib} J.-P.,   {Schaerer} D.,  2018, \mn@doi [\mnras] {10.1093/mnras/sty1820}, \href {https://ui.adsabs.harvard.edu/abs/2018MNRAS.479.5184A} {479, 5184}

\bibitem[\protect\citeauthoryear{{Barrufet} et~al.,}{{Barrufet} et~al.}{2023}]{barrufet2023}
{Barrufet} L.,  et~al., 2023, \mn@doi [\mnras] {10.1093/mnras/stad947}, \href {https://ui.adsabs.harvard.edu/abs/2023MNRAS.522..449B} {522, 449}

\bibitem[\protect\citeauthoryear{{Becker} et~al.,}{{Becker} et~al.}{2001}]{Becker2001}
{Becker} R.~H.,  et~al., 2001, \mn@doi [\aj] {10.1086/324231}, \href {https://ui.adsabs.harvard.edu/abs/2001AJ....122.2850B} {122, 2850}

\bibitem[\protect\citeauthoryear{{Bian}, {Fan}, {McGreer}, {Cai}  \& {Jiang}}{{Bian} et~al.}{2017}]{Bian2017}
{Bian} F.,  {Fan} X.,  {McGreer} I.,  {Cai} Z.,   {Jiang} L.,  2017, \mn@doi [\apjl] {10.3847/2041-8213/aa5ff7}, \href {https://ui.adsabs.harvard.edu/abs/2017ApJ...837L..12B} {837, L12}

\bibitem[\protect\citeauthoryear{{Borthakur}, {Heckman}, {Leitherer}  \& {Overzier}}{{Borthakur} et~al.}{2014}]{Borthakur2014}
{Borthakur} S.,  {Heckman} T.~M.,  {Leitherer} C.,   {Overzier} R.~A.,  2014, \mn@doi [Science] {10.1126/science.1254214}, \href {https://ui.adsabs.harvard.edu/abs/2014Sci...346..216B} {346, 216}

\bibitem[\protect\citeauthoryear{{Bosman} et~al.,}{{Bosman} et~al.}{2022}]{Bosman2022}
{Bosman} S. E.~I.,  et~al., 2022, \mn@doi [\mnras] {10.1093/mnras/stac1046}, \href {https://ui.adsabs.harvard.edu/abs/2022MNRAS.514...55B} {514, 55}

\bibitem[\protect\citeauthoryear{{Bouwens} et~al.,}{{Bouwens} et~al.}{2021}]{Bouwens2021}
{Bouwens} R.~J.,  et~al., 2021, \mn@doi [\aj] {10.3847/1538-3881/abf83e}, \href {https://ui.adsabs.harvard.edu/abs/2021AJ....162...47B} {162, 47}

\bibitem[\protect\citeauthoryear{{Bouwens}, {Illingworth}, {Ellis}, {Oesch}  \& {Stefanon}}{{Bouwens} et~al.}{2022}]{Bouwens2022}
{Bouwens} R.~J.,  {Illingworth} G.~D.,  {Ellis} R.~S.,  {Oesch} P.~A.,   {Stefanon} M.,  2022, arXiv e-prints, \href {https://ui.adsabs.harvard.edu/abs/2022arXiv220511526B} {p. arXiv:2205.11526}

\bibitem[\protect\citeauthoryear{{Bouwens} et~al.,}{{Bouwens} et~al.}{2023}]{Bouwens2023}
{Bouwens} R.~J.,  et~al., 2023, \mn@doi [\mnras] {10.1093/mnras/stad1145}, \href {https://ui.adsabs.harvard.edu/abs/2023MNRAS.523.1036B} {523, 1036}

\bibitem[\protect\citeauthoryear{{Chisholm} et~al.,}{{Chisholm} et~al.}{2022}]{Chisholm2022}
{Chisholm} J.,  et~al., 2022, arXiv e-prints, \href {https://ui.adsabs.harvard.edu/abs/2022arXiv220705771C} {p. arXiv:2207.05771}

\bibitem[\protect\citeauthoryear{{Cowie} \& {Songaila}}{{Cowie} \& {Songaila}}{1986}]{Cowie1986}
{Cowie} L.~L.,  {Songaila} A.,  1986, \mn@doi [\araa] {10.1146/annurev.aa.24.090186.002435}, \href {https://ui.adsabs.harvard.edu/abs/1986ARA&A..24..499C} {24, 499}

\bibitem[\protect\citeauthoryear{{Craig} et~al.,}{{Craig} et~al.}{1997}]{Craig1997}
{Craig} N.,  et~al., 1997, \mn@doi [\apjs] {10.1086/313052}, \href {https://ui.adsabs.harvard.edu/abs/1997ApJS..113..131C} {113, 131}

\bibitem[\protect\citeauthoryear{{Dopita}, {Krauss}, {Sutherland}, {Kobayashi}  \& {Lineweaver}}{{Dopita} et~al.}{2011}]{Dopita2011}
{Dopita} M.~A.,  {Krauss} L.~M.,  {Sutherland} R.~S.,  {Kobayashi} C.,   {Lineweaver} C.~H.,  2011, \mn@doi [\apss] {10.1007/s10509-011-0786-7}, \href {https://ui.adsabs.harvard.edu/abs/2011Ap&SS.335..345D} {335, 345}

\bibitem[\protect\citeauthoryear{{Eide}, {Ciardi}, {Graziani}, {Busch}, {Feng}  \& {Di Matteo}}{{Eide} et~al.}{2020}]{Eide2020}
{Eide} M.~B.,  {Ciardi} B.,  {Graziani} L.,  {Busch} P.,  {Feng} Y.,   {Di Matteo} T.,  2020, \mn@doi [\mnras] {10.1093/mnras/staa2774}, \href {https://ui.adsabs.harvard.edu/abs/2020MNRAS.498.6083E} {498, 6083}

\bibitem[\protect\citeauthoryear{{Eldridge}, {Stanway}, {Xiao}, {McClelland}, {Taylor}, {Ng}, {Greis}  \& {Bray}}{{Eldridge} et~al.}{2017}]{Eldridge2017}
{Eldridge} J.~J.,  {Stanway} E.~R.,  {Xiao} L.,  {McClelland} L.~A.~S.,  {Taylor} G.,  {Ng} M.,  {Greis} S.~M.~L.,   {Bray} J.~C.,  2017, \mn@doi [\pasa] {10.1017/pasa.2017.51}, \href {https://ui.adsabs.harvard.edu/abs/2017PASA...34...58E} {34, e058}

\bibitem[\protect\citeauthoryear{Erickson, Green, Nell, Fleming, Witt, France  \& Stocke}{Erickson et~al.}{2021}]{Erickson2021}
Erickson N.,  Green J.,  Nell N.,  Fleming B.,  Witt E.,  France K.,   Stocke J.,  2021, \mn@doi [Journal of Astronomical Telescopes, Instruments, and Systems] {10.1117/1.JATIS.7.1.015002}, 7, 015002

\bibitem[\protect\citeauthoryear{{Fan}, {Carilli}  \& {Keating}}{{Fan} et~al.}{2006}]{Fan2006}
{Fan} X.,  {Carilli} C.~L.,   {Keating} B.,  2006, \mn@doi [\araa] {10.1146/annurev.astro.44.051905.092514}, \href {https://ui.adsabs.harvard.edu/abs/2006ARA&A..44..415F} {44, 415}

\bibitem[\protect\citeauthoryear{{Ferland} et~al.,}{{Ferland} et~al.}{2017}]{Ferland2017}
{Ferland} G.~J.,  et~al., 2017, \rmxaa, \href {https://ui.adsabs.harvard.edu/abs/2017RMxAA..53..385F} {53, 385}

\bibitem[\protect\citeauthoryear{{Finkelstein} et~al.,}{{Finkelstein} et~al.}{2019}]{Finkelstein2019}
{Finkelstein} S.~L.,  et~al., 2019, \mn@doi [\apj] {10.3847/1538-4357/ab1ea8}, \href {https://ui.adsabs.harvard.edu/abs/2019ApJ...879...36F} {879, 36}

\bibitem[\protect\citeauthoryear{{Fletcher}, {Tang}, {Robertson}, {Nakajima}, {Ellis}, {Stark}  \& {Inoue}}{{Fletcher} et~al.}{2019}]{Fletcher2019}
{Fletcher} T.~J.,  {Tang} M.,  {Robertson} B.~E.,  {Nakajima} K.,  {Ellis} R.~S.,  {Stark} D.~P.,   {Inoue} A.,  2019, \mn@doi [\apj] {10.3847/1538-4357/ab2045}, \href {https://ui.adsabs.harvard.edu/abs/2019ApJ...878...87F} {878, 87}

\bibitem[\protect\citeauthoryear{{Flury} et~al.,}{{Flury} et~al.}{2022a}]{Flury2022a}
{Flury} S.~R.,  et~al., 2022a, \mn@doi [\apjs] {10.3847/1538-4365/ac5331}, \href {https://ui.adsabs.harvard.edu/abs/2022ApJS..260....1F} {260, 1}

\bibitem[\protect\citeauthoryear{{Flury} et~al.,}{{Flury} et~al.}{2022b}]{Flury2022b}
{Flury} S.~R.,  et~al., 2022b, \mn@doi [\apj] {10.3847/1538-4357/ac61e4}, \href {https://ui.adsabs.harvard.edu/abs/2022ApJ...930..126F} {930, 126}

\bibitem[\protect\citeauthoryear{{Harikane}, {Nakajima}, {Ouchi}, {Umeda}, {Isobe}, {Ono}, {Xu}  \& {Zhang}}{{Harikane} et~al.}{2024}]{Harikane2024}
{Harikane} Y.,  {Nakajima} K.,  {Ouchi} M.,  {Umeda} H.,  {Isobe} Y.,  {Ono} Y.,  {Xu} Y.,   {Zhang} Y.,  2024, \mn@doi [\apj] {10.3847/1538-4357/ad0b7e}, \href {https://ui.adsabs.harvard.edu/abs/2024ApJ...960...56H} {960, 56}

\bibitem[\protect\citeauthoryear{{Hassan}, {Dav{\'e}}, {Mitra}, {Finlator}, {Ciardi}  \& {Santos}}{{Hassan} et~al.}{2018}]{Hassan2018}
{Hassan} S.,  {Dav{\'e}} R.,  {Mitra} S.,  {Finlator} K.,  {Ciardi} B.,   {Santos} M.~G.,  2018, \mn@doi [\mnras] {10.1093/mnras/stx2194}, \href {https://ui.adsabs.harvard.edu/abs/2018MNRAS.473..227H} {473, 227}

\bibitem[\protect\citeauthoryear{{Inoue}}{{Inoue}}{2001}]{Inoue2001AJ}
{Inoue} A.~K.,  2001, \mn@doi [\aj] {10.1086/323095}, \href {https://ui.adsabs.harvard.edu/abs/2001AJ....122.1788I} {122, 1788}

\bibitem[\protect\citeauthoryear{{Inoue}}{{Inoue}}{2010}]{Inoue2010}
{Inoue} A.~K.,  2010, \mn@doi [\mnras] {10.1111/j.1365-2966.2009.15730.x}, \href {https://ui.adsabs.harvard.edu/abs/2010MNRAS.401.1325I} {401, 1325}

\bibitem[\protect\citeauthoryear{{Inoue}, {Hirashita}  \& {Kamaya}}{{Inoue} et~al.}{2001}]{Inoue2001ApJ}
{Inoue} A.~K.,  {Hirashita} H.,   {Kamaya} H.,  2001, \mn@doi [\apj] {10.1086/321499}, \href {https://ui.adsabs.harvard.edu/abs/2001ApJ...555..613I} {555, 613}

\bibitem[\protect\citeauthoryear{{Inoue}, {Iwata}  \& {Deharveng}}{{Inoue} et~al.}{2006}]{Inoue2006}
{Inoue} A.~K.,  {Iwata} I.,   {Deharveng} J.-M.,  2006, \mn@doi [\mnras] {10.1111/j.1745-3933.2006.00195.x}, \href {https://ui.adsabs.harvard.edu/abs/2006MNRAS.371L...1I} {371, L1}

\bibitem[\protect\citeauthoryear{{Inoue} et~al.,}{{Inoue} et~al.}{2011}]{Inoue2011}
{Inoue} A.~K.,  et~al., 2011, \mn@doi [\mnras] {10.1111/j.1365-2966.2010.17851.x}, \href {https://ui.adsabs.harvard.edu/abs/2011MNRAS.411.2336I} {411, 2336}

\bibitem[\protect\citeauthoryear{{Inoue}, {Shimizu}, {Iwata}  \& {Tanaka}}{{Inoue} et~al.}{2014}]{Inoue2014}
{Inoue} A.~K.,  {Shimizu} I.,  {Iwata} I.,   {Tanaka} M.,  2014, \mn@doi [\mnras] {10.1093/mnras/stu936}, \href {https://ui.adsabs.harvard.edu/abs/2014MNRAS.442.1805I} {442, 1805}

\bibitem[\protect\citeauthoryear{{Izotov}, {Schaerer}, {Thuan}, {Worseck}, {Guseva}, {Orlitov{\'a}}  \& {Verhamme}}{{Izotov} et~al.}{2016a}]{Izotov2016b}
{Izotov} Y.~I.,  {Schaerer} D.,  {Thuan} T.~X.,  {Worseck} G.,  {Guseva} N.~G.,  {Orlitov{\'a}} I.,   {Verhamme} A.,  2016a, \mn@doi [\mnras] {10.1093/mnras/stw1205}, \href {https://ui.adsabs.harvard.edu/abs/2016MNRAS.461.3683I} {461, 3683}

\bibitem[\protect\citeauthoryear{{Izotov}, {Orlitov{\'a}}, {Schaerer}, {Thuan}, {Verhamme}, {Guseva}  \& {Worseck}}{{Izotov} et~al.}{2016b}]{Izotov2016a}
{Izotov} Y.~I.,  {Orlitov{\'a}} I.,  {Schaerer} D.,  {Thuan} T.~X.,  {Verhamme} A.,  {Guseva} N.~G.,   {Worseck} G.,  2016b, \mn@doi [\nat] {10.1038/nature16456}, \href {https://ui.adsabs.harvard.edu/abs/2016Natur.529..178I} {529, 178}

\bibitem[\protect\citeauthoryear{{Izotov}, {Schaerer}, {Worseck}, {Guseva}, {Thuan}, {Verhamme}, {Orlitov{\'a}}  \& {Fricke}}{{Izotov} et~al.}{2018a}]{Izotov2018a}
{Izotov} Y.~I.,  {Schaerer} D.,  {Worseck} G.,  {Guseva} N.~G.,  {Thuan} T.~X.,  {Verhamme} A.,  {Orlitov{\'a}} I.,   {Fricke} K.~J.,  2018a, \mn@doi [\mnras] {10.1093/mnras/stx3115}, \href {https://ui.adsabs.harvard.edu/abs/2018MNRAS.474.4514I} {474, 4514}

\bibitem[\protect\citeauthoryear{{Izotov}, {Worseck}, {Schaerer}, {Guseva}, {Thuan}, {Fricke}  \& {Orlitov{\'a}}}{{Izotov} et~al.}{2018b}]{Izotov2018b}
{Izotov} Y.~I.,  {Worseck} G.,  {Schaerer} D.,  {Guseva} N.~G.,  {Thuan} T.~X.,  {Fricke} Verhamme A.,   {Orlitov{\'a}} I.,  2018b, \mn@doi [\mnras] {10.1093/mnras/sty1378}, \href {https://ui.adsabs.harvard.edu/abs/2018MNRAS.478.4851I} {478, 4851}

\bibitem[\protect\citeauthoryear{{Izotov}, {Worseck}, {Schaerer}, {Guseva}, {Chisholm}, {Thuan}, {Fricke}  \& {Verhamme}}{{Izotov} et~al.}{2021}]{Izotov2021}
{Izotov} Y.~I.,  {Worseck} G.,  {Schaerer} D.,  {Guseva} N.~G.,  {Chisholm} J.,  {Thuan} T.~X.,  {Fricke} K.~J.,   {Verhamme} A.,  2021, \mn@doi [\mnras] {10.1093/mnras/stab612}, \href {https://ui.adsabs.harvard.edu/abs/2021MNRAS.503.1734I} {503, 1734}

\bibitem[\protect\citeauthoryear{{Ji}, {Yan}, {Riffel}, {Drory}  \& {Zhang}}{{Ji} et~al.}{2020}]{Ji2020}
{Ji} X.,  {Yan} R.,  {Riffel} R.,  {Drory} N.,   {Zhang} K.,  2020, \mn@doi [\mnras] {10.1093/mnras/staa1521}, \href {https://ui.adsabs.harvard.edu/abs/2020MNRAS.496.1262J} {496, 1262}

\bibitem[\protect\citeauthoryear{{Johnson} \& {Khochfar}}{{Johnson} \& {Khochfar}}{2011}]{Johnson2011}
{Johnson} J.~L.,  {Khochfar} S.,  2011, \mn@doi [\apj] {10.1088/0004-637X/743/2/126}, \href {https://ui.adsabs.harvard.edu/abs/2011ApJ...743..126J} {743, 126}

\bibitem[\protect\citeauthoryear{{Katz} et~al.,}{{Katz} et~al.}{2023}]{Katz2023}
{Katz} H.,  et~al., 2023, \mn@doi [The Open Journal of Astrophysics] {10.21105/astro.2309.03269}, \href {https://ui.adsabs.harvard.edu/abs/2023OJAp....6E..44K} {6, 44}

\bibitem[\protect\citeauthoryear{{Kerutt} et~al.,}{{Kerutt} et~al.}{2023}]{Kerutt2023}
{Kerutt} J.,  et~al., 2023, arXiv e-prints, \href {https://ui.adsabs.harvard.edu/abs/2023arXiv231208791K} {p. arXiv:2312.08791}

\bibitem[\protect\citeauthoryear{{Kulkarni}, {Keating}, {Haehnelt}, {Bosman}, {Puchwein}, {Chardin}  \& {Aubert}}{{Kulkarni} et~al.}{2019}]{Kulkarni2019}
{Kulkarni} G.,  {Keating} L.~C.,  {Haehnelt} M.~G.,  {Bosman} S. E.~I.,  {Puchwein} E.,  {Chardin} J.,   {Aubert} D.,  2019, \mn@doi [\mnras] {10.1093/mnrasl/slz025}, \href {https://ui.adsabs.harvard.edu/abs/2019MNRAS.485L..24K} {485, L24}

\bibitem[\protect\citeauthoryear{{Leitet}, {Bergvall}, {Hayes}, {Linn{\'e}}  \& {Zackrisson}}{{Leitet} et~al.}{2013}]{Leitet2013}
{Leitet} E.,  {Bergvall} N.,  {Hayes} M.,  {Linn{\'e}} S.,   {Zackrisson} E.,  2013, \mn@doi [\aap] {10.1051/0004-6361/201118370}, \href {https://ui.adsabs.harvard.edu/abs/2013A&A...553A.106L} {553, A106}

\bibitem[\protect\citeauthoryear{{Leitherer} et~al.,}{{Leitherer} et~al.}{1999}]{Leitherer1999}
{Leitherer} C.,  et~al., 1999, \mn@doi [\apjs] {10.1086/313233}, \href {https://ui.adsabs.harvard.edu/abs/1999ApJS..123....3L} {123, 3}

\bibitem[\protect\citeauthoryear{{Leitherer}, {Hernandez}, {Lee}  \& {Oey}}{{Leitherer} et~al.}{2016}]{Leitherer2016}
{Leitherer} C.,  {Hernandez} S.,  {Lee} J.~C.,   {Oey} M.~S.,  2016, \mn@doi [\apj] {10.3847/0004-637X/823/1/64}, \href {https://ui.adsabs.harvard.edu/abs/2016ApJ...823...64L} {823, 64}

\bibitem[\protect\citeauthoryear{{Lewis}, {Ocvirk}, {Dubois}, {Aubert}, {Chardin}, {Gillet}  \& {Th{\'e}lie}}{{Lewis} et~al.}{2023}]{Lewis2023}
{Lewis} J. S.~W.,  {Ocvirk} P.,  {Dubois} Y.,  {Aubert} D.,  {Chardin} J.,  {Gillet} N.,   {Th{\'e}lie} {\'E}.,  2023, \mn@doi [\mnras] {10.1093/mnras/stad081}, \href {https://ui.adsabs.harvard.edu/abs/2023MNRAS.519.5987L} {519, 5987}

\bibitem[\protect\citeauthoryear{{Ma}, {Quataert}, {Wetzel}, {Hopkins}, {Faucher-Gigu{\`e}re}  \& {Kere{\v{s}}}}{{Ma} et~al.}{2020}]{Ma2020}
{Ma} X.,  {Quataert} E.,  {Wetzel} A.,  {Hopkins} P.~F.,  {Faucher-Gigu{\`e}re} C.-A.,   {Kere{\v{s}}} D.,  2020, \mn@doi [\mnras] {10.1093/mnras/staa2404}, \href {https://ui.adsabs.harvard.edu/abs/2020MNRAS.498.2001M} {498, 2001}

\bibitem[\protect\citeauthoryear{{Madau}}{{Madau}}{1995}]{Madau1995}
{Madau} P.,  1995, \mn@doi [\apj] {10.1086/175332}, \href {https://ui.adsabs.harvard.edu/abs/1995ApJ...441...18M} {441, 18}

\bibitem[\protect\citeauthoryear{{Madau} \& {Haardt}}{{Madau} \& {Haardt}}{2015}]{Madau2015}
{Madau} P.,  {Haardt} F.,  2015, \mn@doi [\apjl] {10.1088/2041-8205/813/1/L8}, \href {https://ui.adsabs.harvard.edu/abs/2015ApJ...813L...8M} {813, L8}

\bibitem[\protect\citeauthoryear{{Madau}, {Haardt}  \& {Rees}}{{Madau} et~al.}{1999}]{Madau1999}
{Madau} P.,  {Haardt} F.,   {Rees} M.~J.,  1999, \mn@doi [\apj] {10.1086/306975}, \href {https://ui.adsabs.harvard.edu/abs/1999ApJ...514..648M} {514, 648}

\bibitem[\protect\citeauthoryear{{Maji} et~al.,}{{Maji} et~al.}{2022}]{Maji2022}
{Maji} M.,  et~al., 2022, \mn@doi [\aap] {10.1051/0004-6361/202142740}, \href {https://ui.adsabs.harvard.edu/abs/2022A&A...663A..66M} {663, A66}

\bibitem[\protect\citeauthoryear{{Marconi} et~al.,}{{Marconi} et~al.}{2024}]{Marconi2024}
{Marconi} A.,  et~al., 2024, \mn@doi [arXiv e-prints] {10.48550/arXiv.2401.13028}, \href {https://ui.adsabs.harvard.edu/abs/2024arXiv240113028M} {p. arXiv:2401.13028}

\bibitem[\protect\citeauthoryear{{Marques-Chaves}, {Schaerer}, {{\'A}lvarez-M{\'a}rquez}, {Colina}, {Dessauges-Zavadsky}, {P{\'e}rez-Fournon}, {Saldana-Lopez}  \& {Verhamme}}{{Marques-Chaves} et~al.}{2021}]{Marques-Chaves2021}
{Marques-Chaves} R.,  {Schaerer} D.,  {{\'A}lvarez-M{\'a}rquez} J.,  {Colina} L.,  {Dessauges-Zavadsky} M.,  {P{\'e}rez-Fournon} I.,  {Saldana-Lopez} A.,   {Verhamme} A.,  2021, \mn@doi [\mnras] {10.1093/mnras/stab2187}, \href {https://ui.adsabs.harvard.edu/abs/2021MNRAS.507..524M} {507, 524}

\bibitem[\protect\citeauthoryear{{Marques-Chaves} et~al.,}{{Marques-Chaves} et~al.}{2022}]{Marques-Chaves2022}
{Marques-Chaves} R.,  et~al., 2022, \mn@doi [\mnras] {10.1093/mnras/stac2893}, \href {https://ui.adsabs.harvard.edu/abs/2022MNRAS.517.2972M} {517, 2972}

\bibitem[\protect\citeauthoryear{{Mauerhofer}, {Verhamme}, {Blaizot}, {Garel}, {Kimm}, {Michel-Dansac}  \& {Rosdahl}}{{Mauerhofer} et~al.}{2021}]{Mauerhofer2021}
{Mauerhofer} V.,  {Verhamme} A.,  {Blaizot} J.,  {Garel} T.,  {Kimm} T.,  {Michel-Dansac} L.,   {Rosdahl} J.,  2021, \mn@doi [\aap] {10.1051/0004-6361/202039449}, \href {https://ui.adsabs.harvard.edu/abs/2021A&A...646A..80M} {646, A80}

\bibitem[\protect\citeauthoryear{{Mirabel}, {Dijkstra}, {Laurent}, {Loeb}  \& {Pritchard}}{{Mirabel} et~al.}{2011}]{Mirabel2011}
{Mirabel} I.~F.,  {Dijkstra} M.,  {Laurent} P.,  {Loeb} A.,   {Pritchard} J.~R.,  2011, \mn@doi [\aap] {10.1051/0004-6361/201016357}, \href {https://ui.adsabs.harvard.edu/abs/2011A&A...528A.149M} {528, A149}

\bibitem[\protect\citeauthoryear{Mitsuhashi et~al.,}{Mitsuhashi et~al.}{2023}]{Mitsuhashi2023}
Mitsuhashi I.,  et~al., 2023, The ALMA-CRISTAL survey: Widespread dust-obscured star formation in typical star-forming galaxies at z=4-6 (\mn@eprint {arXiv} {2311.17671})

\bibitem[\protect\citeauthoryear{{Mostardi}, {Shapley}, {Steidel}, {Trainor}, {Reddy}  \& {Siana}}{{Mostardi} et~al.}{2015}]{Mostardi2015}
{Mostardi} R.~E.,  {Shapley} A.~E.,  {Steidel} C.~C.,  {Trainor} R.~F.,  {Reddy} N.~A.,   {Siana} B.,  2015, \mn@doi [\apj] {10.1088/0004-637X/810/2/107}, \href {https://ui.adsabs.harvard.edu/abs/2015ApJ...810..107M} {810, 107}

\bibitem[\protect\citeauthoryear{{Nelson} et~al.,}{{Nelson} et~al.}{2023}]{Nelson2023}
{Nelson} E.~J.,  et~al., 2023, \mn@doi [\apjl] {10.3847/2041-8213/acc1e1}, \href {https://ui.adsabs.harvard.edu/abs/2023ApJ...948L..18N} {948, L18}

\bibitem[\protect\citeauthoryear{{Onoue} et~al.,}{{Onoue} et~al.}{2017}]{Onoue2017}
{Onoue} M.,  et~al., 2017, \mn@doi [\apjl] {10.3847/2041-8213/aa8cc6}, \href {https://ui.adsabs.harvard.edu/abs/2017ApJ...847L..15O} {847, L15}

\bibitem[\protect\citeauthoryear{{Paardekooper}, {Khochfar}  \& {Dalla Vecchia}}{{Paardekooper} et~al.}{2015}]{Paardekooper2015}
{Paardekooper} J.-P.,  {Khochfar} S.,   {Dalla Vecchia} C.,  2015, \mn@doi [\mnras] {10.1093/mnras/stv1114}, \href {https://ui.adsabs.harvard.edu/abs/2015MNRAS.451.2544P} {451, 2544}

\bibitem[\protect\citeauthoryear{{Prieto-Lyon} et~al.,}{{Prieto-Lyon} et~al.}{2023}]{Prieto-Lyon2023}
{Prieto-Lyon} G.,  et~al., 2023, \mn@doi [\aap] {10.1051/0004-6361/202245532}, \href {https://ui.adsabs.harvard.edu/abs/2023A&A...672A.186P} {672, A186}

\bibitem[\protect\citeauthoryear{{Rinaldi} et~al.,}{{Rinaldi} et~al.}{2023}]{Rinaldi2023}
{Rinaldi} P.,  et~al., 2023, \mn@doi [\apj] {10.3847/1538-4357/acdc27}, \href {https://ui.adsabs.harvard.edu/abs/2023ApJ...952..143R} {952, 143}

\bibitem[\protect\citeauthoryear{{Rivera-Thorsen} et~al.,}{{Rivera-Thorsen} et~al.}{2019}]{Rivera-Thorsen2019}
{Rivera-Thorsen} T.~E.,  et~al., 2019, \mn@doi [Science] {10.1126/science.aaw0978}, \href {https://ui.adsabs.harvard.edu/abs/2019Sci...366..738R} {366, 738}

\bibitem[\protect\citeauthoryear{{Robertson} et~al.,}{{Robertson} et~al.}{2013}]{Robertson2013}
{Robertson} B.~E.,  et~al., 2013, \mn@doi [\apj] {10.1088/0004-637X/768/1/71}, \href {https://ui.adsabs.harvard.edu/abs/2013ApJ...768...71R} {768, 71}

\bibitem[\protect\citeauthoryear{{Robertson}, {Ellis}, {Furlanetto}  \& {Dunlop}}{{Robertson} et~al.}{2015}]{Robertson2015}
{Robertson} B.~E.,  {Ellis} R.~S.,  {Furlanetto} S.~R.,   {Dunlop} J.~S.,  2015, \mn@doi [\apjl] {10.1088/2041-8205/802/2/L19}, \href {https://ui.adsabs.harvard.edu/abs/2015ApJ...802L..19R} {802, L19}

\bibitem[\protect\citeauthoryear{{Rosdahl} et~al.,}{{Rosdahl} et~al.}{2018}]{Rosdahl2018}
{Rosdahl} J.,  et~al., 2018, \mn@doi [\mnras] {10.1093/mnras/sty1655}, \href {https://ui.adsabs.harvard.edu/abs/2018MNRAS.479..994R} {479, 994}

\bibitem[\protect\citeauthoryear{{Rosdahl} et~al.,}{{Rosdahl} et~al.}{2022}]{Rosdahl2022}
{Rosdahl} J.,  et~al., 2022, arXiv e-prints, \href {https://ui.adsabs.harvard.edu/abs/2022arXiv220703232R} {p. arXiv:2207.03232}

\bibitem[\protect\citeauthoryear{{Saha} et~al.,}{{Saha} et~al.}{2020}]{Saha2020}
{Saha} K.,  et~al., 2020, \mn@doi [Nature Astronomy] {10.1038/s41550-020-1173-5}, \href {https://ui.adsabs.harvard.edu/abs/2020NatAs...4.1185S} {4, 1185}

\bibitem[\protect\citeauthoryear{{Savage} \& {Sembach}}{{Savage} \& {Sembach}}{1996}]{Savage1996}
{Savage} B.~D.,  {Sembach} K.~R.,  1996, \mn@doi [\araa] {10.1146/annurev.astro.34.1.279}, \href {https://ui.adsabs.harvard.edu/abs/1996ARA&A..34..279S} {34, 279}

\bibitem[\protect\citeauthoryear{{Schaerer}}{{Schaerer}}{2003}]{Schaerer2003}
{Schaerer} D.,  2003, \mn@doi [\aap] {10.1051/0004-6361:20021525}, \href {https://ui.adsabs.harvard.edu/abs/2003A&A...397..527S} {397, 527}

\bibitem[\protect\citeauthoryear{{Seeyave} et~al.,}{{Seeyave} et~al.}{2023}]{Seeyave2023}
{Seeyave} L. T.~C.,  et~al., 2023, \mn@doi [\mnras] {10.1093/mnras/stad2487}, \href {https://ui.adsabs.harvard.edu/abs/2023MNRAS.525.2422S} {525, 2422}

\bibitem[\protect\citeauthoryear{{Simmonds}, {Schaerer}  \& {Verhamme}}{{Simmonds} et~al.}{2021}]{Simmonds2021}
{Simmonds} C.,  {Schaerer} D.,   {Verhamme} A.,  2021, \mn@doi [\aap] {10.1051/0004-6361/202141856}, \href {https://ui.adsabs.harvard.edu/abs/2021A&A...656A.127S} {656, A127}

\bibitem[\protect\citeauthoryear{{Simmonds} et~al.,}{{Simmonds} et~al.}{2023}]{Simmonds2023b}
{Simmonds} C.,  et~al., 2023, \mn@doi [arXiv e-prints] {10.48550/arXiv.2310.01112}, \href {https://ui.adsabs.harvard.edu/abs/2023arXiv231001112S} {p. arXiv:2310.01112}

\bibitem[\protect\citeauthoryear{{Stefanon}, {Bouwens}, {Illingworth}, {Labb{\'e}}, {Oesch}  \& {Gonzalez}}{{Stefanon} et~al.}{2022}]{Stefanon2022}
{Stefanon} M.,  {Bouwens} R.~J.,  {Illingworth} G.~D.,  {Labb{\'e}} I.,  {Oesch} P.~A.,   {Gonzalez} V.,  2022, \mn@doi [\apj] {10.3847/1538-4357/ac7e44}, \href {https://ui.adsabs.harvard.edu/abs/2022ApJ...935...94S} {935, 94}

\bibitem[\protect\citeauthoryear{Steidel, Bogosavljevi{\'c}, Shapley, Reddy, Rudie, Pettini, Trainor  \& Strom}{Steidel et~al.}{2018}]{Steidel2018}
Steidel C.~C.,  Bogosavljevi{\'c} M.,  Shapley A.~E.,  Reddy N.~A.,  Rudie G.~C.,  Pettini M.,  Trainor R.~F.,   Strom A.~L.,  2018, The Astrophysical Journal, 869, 123

\bibitem[\protect\citeauthoryear{{Thai} et~al.,}{{Thai} et~al.}{2023}]{Thai2023}
{Thai} T.~T.,  et~al., 2023, \mn@doi [\aap] {10.1051/0004-6361/202346716}, \href {https://ui.adsabs.harvard.edu/abs/2023A&A...678A.139T} {678, A139}

\bibitem[\protect\citeauthoryear{{Trebitsch}, {Volonteri}  \& {Dubois}}{{Trebitsch} et~al.}{2020}]{Trebitsch2020}
{Trebitsch} M.,  {Volonteri} M.,   {Dubois} Y.,  2020, \mn@doi [\mnras] {10.1093/mnras/staa1012}, \href {https://ui.adsabs.harvard.edu/abs/2020MNRAS.494.3453T} {494, 3453}

\bibitem[\protect\citeauthoryear{{Trebitsch} et~al.,}{{Trebitsch} et~al.}{2021}]{Trebitsch2021}
{Trebitsch} M.,  et~al., 2021, \mn@doi [\aap] {10.1051/0004-6361/202037698}, \href {https://ui.adsabs.harvard.edu/abs/2021A&A...653A.154T} {653, A154}

\bibitem[\protect\citeauthoryear{{Trebitsch}, {Hutter}, {Dayal}, {Gottl{\"o}ber}, {Legrand}  \& {Yepes}}{{Trebitsch} et~al.}{2023}]{Trebitsch2023}
{Trebitsch} M.,  {Hutter} A.,  {Dayal} P.,  {Gottl{\"o}ber} S.,  {Legrand} L.,   {Yepes} G.,  2023, \mn@doi [\mnras] {10.1093/mnras/stac2138}, \href {https://ui.adsabs.harvard.edu/abs/2023MNRAS.518.3576T} {518, 3576}

\bibitem[\protect\citeauthoryear{{Vanzella} et~al.,}{{Vanzella} et~al.}{2016}]{Vanzella2016}
{Vanzella} E.,  et~al., 2016, \mn@doi [\apj] {10.3847/0004-637X/825/1/41}, \href {https://ui.adsabs.harvard.edu/abs/2016ApJ...825...41V} {825, 41}

\bibitem[\protect\citeauthoryear{{Vanzella} et~al.,}{{Vanzella} et~al.}{2018}]{Vanzella2018}
{Vanzella} E.,  et~al., 2018, \mn@doi [\mnras] {10.1093/mnrasl/sly023}, \href {https://ui.adsabs.harvard.edu/abs/2018MNRAS.476L..15V} {476, L15}

\bibitem[\protect\citeauthoryear{{Wang}, {Heckman}, {Leitherer}, {Alexandroff}, {Borthakur}  \& {Overzier}}{{Wang} et~al.}{2019}]{Wang2019}
{Wang} B.,  {Heckman} T.~M.,  {Leitherer} C.,  {Alexandroff} R.,  {Borthakur} S.,   {Overzier} R.~A.,  2019, \mn@doi [\apj] {10.3847/1538-4357/ab418f}, \href {https://ui.adsabs.harvard.edu/abs/2019ApJ...885...57W} {885, 57}

\bibitem[\protect\citeauthoryear{{Xu}, {Wise}, {Norman}, {Ahn}  \& {O'Shea}}{{Xu} et~al.}{2016}]{Xu2016}
{Xu} H.,  {Wise} J.~H.,  {Norman} M.~L.,  {Ahn} K.,   {O'Shea} B.~W.,  2016, \mn@doi [\apj] {10.3847/1538-4357/833/1/84}, \href {https://ui.adsabs.harvard.edu/abs/2016ApJ...833...84X} {833, 84}

\bibitem[\protect\citeauthoryear{{Yajima}, {Choi}  \& {Nagamine}}{{Yajima} et~al.}{2011}]{Yajima2011}
{Yajima} H.,  {Choi} J.-H.,   {Nagamine} K.,  2011, \mn@doi [\mnras] {10.1111/j.1365-2966.2010.17920.x}, \href {https://ui.adsabs.harvard.edu/abs/2011MNRAS.412..411Y} {412, 411}

\bibitem[\protect\citeauthoryear{{Yang} et~al.,}{{Yang} et~al.}{2020}]{Yang2020}
{Yang} J.,  et~al., 2020, \mn@doi [\apj] {10.3847/1538-4357/abbc1b}, \href {https://ui.adsabs.harvard.edu/abs/2020ApJ...904...26Y} {904, 26}

\bibitem[\protect\citeauthoryear{{Yeh} et~al.,}{{Yeh} et~al.}{2023}]{Yeh2023}
{Yeh} J. Y.~C.,  et~al., 2023, \mn@doi [\mnras] {10.1093/mnras/stad210}, \href {https://ui.adsabs.harvard.edu/abs/2023MNRAS.520.2757Y} {520, 2757}

\bibitem[\protect\citeauthoryear{{de Barros} et~al.,}{{de Barros} et~al.}{2016}]{DeBarros2016}
{de Barros} S.,  et~al., 2016, \mn@doi [\aap] {10.1051/0004-6361/201527046}, \href {https://ui.adsabs.harvard.edu/abs/2016A&A...585A..51D} {585, A51}

\makeatother
\end{thebibliography}


\appendix

\section{Starburst99 models}
\label{appendix}
In this Appendix we use grid of \textsc{Cloudy} models using Starburst99 \citep[SB99; ][]{Leitherer1999} stellar populations, to reproduce some of the key figures of the paper. The stellar and nebular parameters used in this section are presented in Table~\ref{appendix}. All the SB99 models were computed for a mass of $10^6$ M$_{\odot}$ stars, formed in a burst. As in the rest of this work, the element abundances/grains are set to the predefined "ism" abundance in \textsc{Cloudy}.\\

\newpage
\begin{table}
        \centering
        \begin{tabular}{cc}
        \hline
        \noalign{\smallskip}
        Parameter & Values \\ 
        \noalign{\smallskip}
        \hline
        \noalign{\smallskip}
            Age [Myr] & 0.5, 1, 5, 10\\ 
            $Z$ [SB99] & 0.0004, 0.008, 0.050\\
            IMF [$\alpha_2$] & -0.8, -1.7, -2.3\\
            log(\nhstop)/[cm$^{-2}$]& $16, 16.5, 17, 17.5, 18, 18.5, 19$\\
            log<U> & $-3.5, -3.0, -2.5, -2.0, -1.5, -1.0$\\
            log($\rho$) [cm$^{-3}$] & 0, 1, 2, 3\\
        \noalign{\smallskip}
        \hline
        \noalign{\smallskip}
        \end{tabular}
        \caption{Properties of the intrinsic SB99 spectra used in the models. All the SB99 stellar populations were computed for a burst of $10^6$ M$_{\odot}$. \textsl{Row 1:} age of the population. \textsl{Row 2:} metallicity of the stellar population, where $Z = 0.020$ corresponds to solar metallicity. \textsl{Row 3:} higher mass ($0.5 <$ M $300$ M$_{\odot}$) IMF exponent, where $-2.3$ is considered canonical. \textsl{Row 4:}  logarithm of the neutral hydrogen column values used as
        stopping criteria in \textsc{Cloudy}. \textsl{Row 5:} ionisation parameters probed in the grid. \textsl{Row 6:} densities explored in this appendix.}
        \label{tab:appendix_parameters}
    \end{table}

\begin{figure}
    \centering
    \includegraphics[width=1\columnwidth]{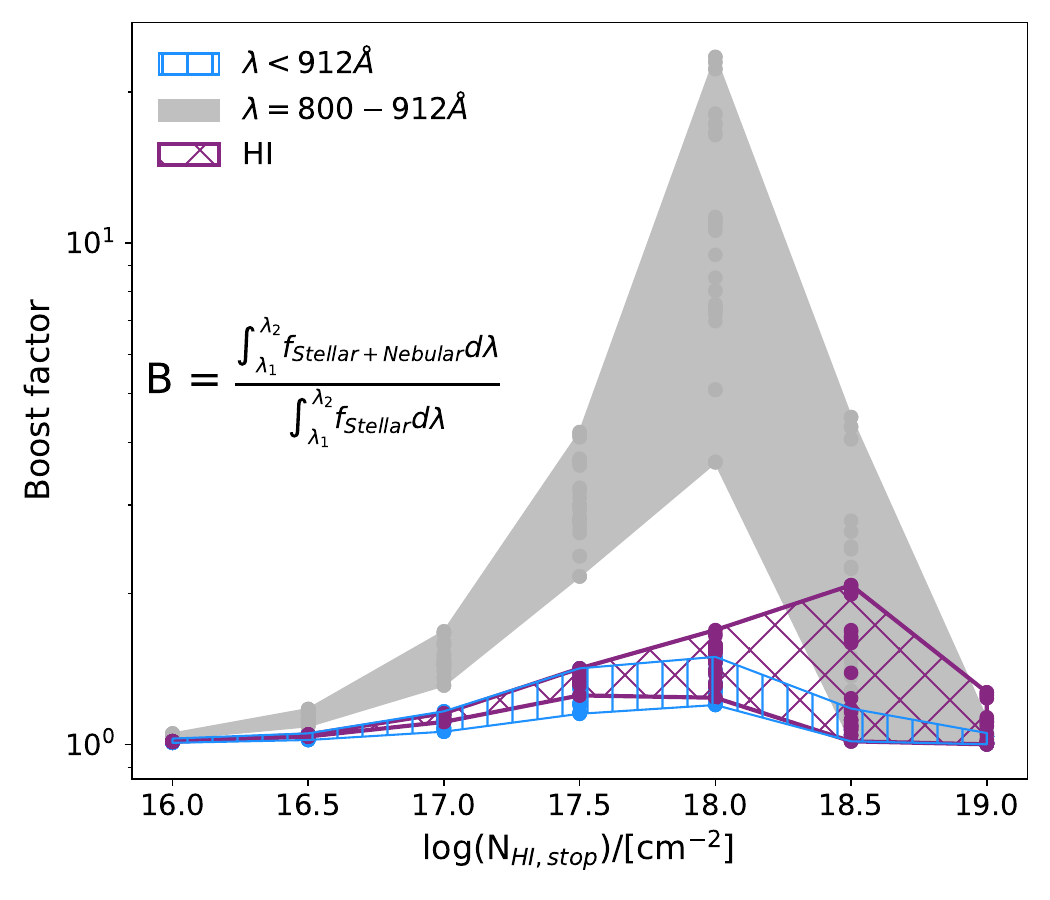}
    \caption{Reproduction of Figure~\ref{fig:boost} but now using all the SB99 models described in Table~\ref{tab:appendix_parameters}. The shaded areas show the strength of the LyC boost as a function of \nhstop\/ for $\lambda < 912$ \AA\/ (blue hatched area), $\lambda = 800-912$ \AA\/ (grey shaded area), and $\lambda = 504-912$ \AA\/ (purple hatched area; "HI"). As with the BPASS models, the strongest boost is found in the spectral region closest to the Lyman limit.}
    \label{fig:app_boost}
\end{figure}

\begin{figure}
    \centering
    \includegraphics[width=0.8\columnwidth]{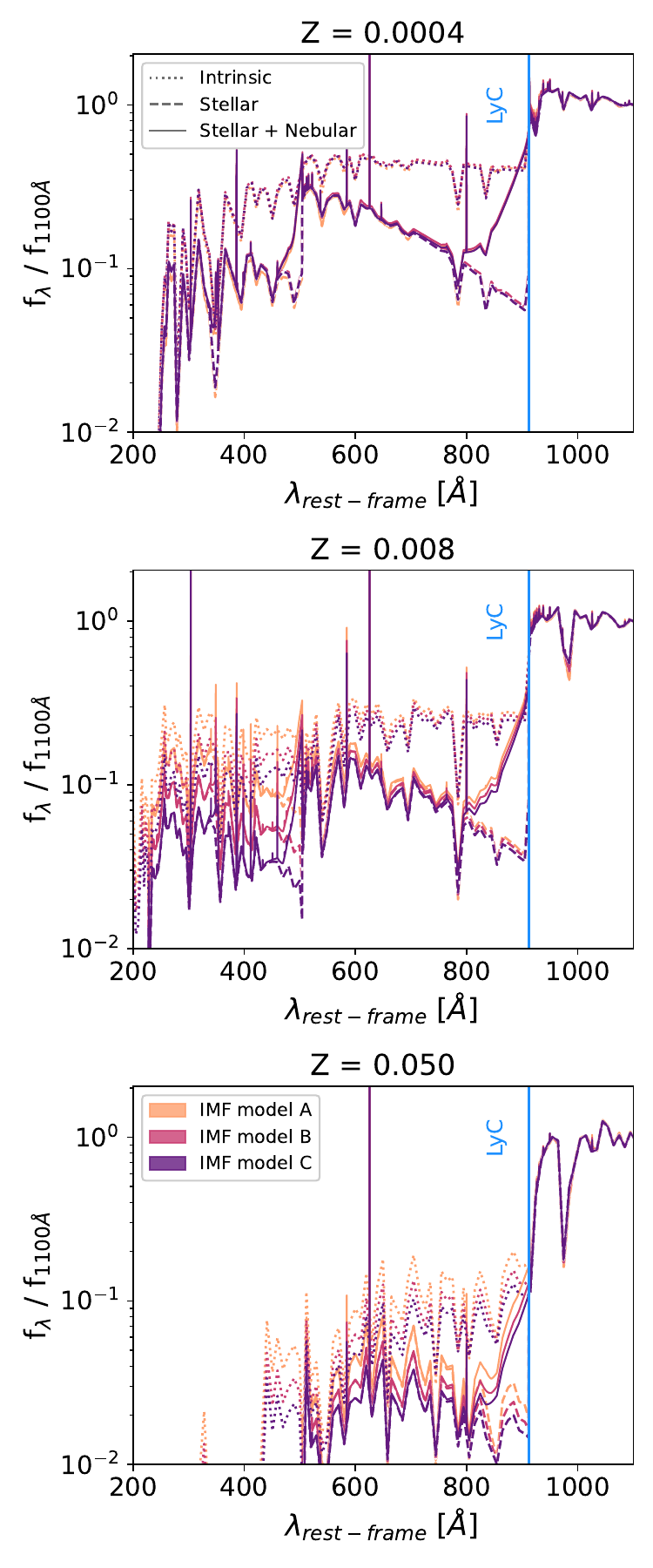}
    \caption{Reproduction of Figure~\ref{fig:intrinsic_LyC} but now using SB99 models. Shape of LyC emission depending on initial mass function (IMF) and metallicity, for a stellar population of age $\sim$5 Myr formed in a burst. Each panel shows the intrinsic (filled lines), stellar (dotted lines) and stellar+nebular (shaded area) flux densities, assuming a \textsc{Cloudy} model with log(\nhstop\/)/[cm$^{-2}$] = 17.5, for a fixed metallicity as indicated in the title, and colour-coded by the IMF model. All values have been normalised to the flux density at $\lambda = 1100$ \AA\/. \textsl{Model A:} $\alpha_2 = -0.8$. \textsl{Model B:} $\alpha_2 = -1.7$. \textsl{Model C:}  $\alpha_2 = -2.3$. Where $\alpha_2$ is the higher mass (0.5 < M < 120 M$_{\odot}$) exponent, assuming a broken power law IMF. As in the case of the BPASS models, it can be seen that for a fixed age and metallicity, the choice of IMF results in only a slight difference in the shape of the ionising spectra, but that the nebular emission produces a flux excess at $\lambda = 800 - 900$ \AA\/ in every case. }
    \label{fig:app_intrinsic_LyC}
\end{figure}

Figure~\ref{fig:app_boost} shows the boost factor as a function of \nhstop\/ for the more relevant wavelength bins studied in this work, while Figure~\ref{fig:app_intrinsic_LyC} shows the dependence of the LyC spectra on metallicity and IMF shape, adopting higher mass exponents of $\alpha_2 = -0.8$, $-1.7$ and $-2.3$.  
We find that all the trends and conclusions are in agreement with those made through the use of BPASS models. The differences in spectral shapes arise mainly due to the star formation history assumed (i.e. burst versus constant).

    \label{fig:app_parameters_bump_effect}


\bsp	
\label{lastpage}
\end{document}